\documentclass[pra,twocolumn,showpacs,showkeys,amsmath,amsfonts,amssymb,nofootinbib]{revtex4}
\usepackage{bm}
\usepackage{bbm}
\usepackage{amssymb}
\usepackage{amsmath}
\usepackage{amsfonts}
\usepackage{graphicx}
\usepackage{psfrag}

\expandafter\ifx\csname package@font\endcsname\relax\else
 \expandafter\expandafter
 \expandafter\usepackage
 \expandafter\expandafter
 \expandafter{\csname package@font\endcsname}
\fi

\newtheorem{define}{Definition}
\newtheorem{theo}{Theorem}
\newtheorem{lemma}[theo]{Lemma}
\newtheorem{propo}[theo]{Proposition}
\newtheorem{coro}[theo]{Corollary}
\newtheorem{remark}{Remark}

\newcommand{\ket}[1]{\ensuremath{|#1 \rangle}}
\newcommand{\bra}[1]{\ensuremath{\langle #1|}}

\newcommand{\kb}[1]{\ensuremath{| #1 \rangle \! \langle #1 |}}

\newcommand{\ip}[2]{\ensuremath{\langle #1 | #2 \rangle}}
\newcommand{\op}[2]{\ensuremath{| #1 \rangle \! \langle #2 |}}

\newcommand{\1}{\protect\ensuremath{\mathbbm{1}}}
\newcommand{\F}{\ensuremath{\mathbb{F}}}

\newcommand{\abs}[1]{\ensuremath{| #1 |}}
\newcommand{\cb}[1]{\ensuremath{\| #1 \|_{cb}}}
\newcommand{\norm}[1]{\ensuremath{\| #1 \|_{\infty}}}
\newcommand{\tracenorm}[1]{\ensuremath{\| #1 \|_{1}}}

\newcommand{\C}{\ensuremath{\mathbb{C}}}
\newcommand{\R}{\ensuremath{\mathbb{R}}}
\newcommand{\N}{\ensuremath{\mathbb{N}}}
\newcommand{\Z}{\ensuremath{\mathbb{Z}}}

\newcommand{\tr}[1]{\ensuremath{{\rm tr}(#1)}}
\newcommand{\ld}{\ensuremath{{\rm ld} \, }}
\newcommand{\id}{\ensuremath{{\rm id} \, }}

\newcommand{\hh}{\ensuremath{\mathcal{H}}}
\newcommand{\bh}{\ensuremath{\mathcal{B(H)}}}
\newcommand{\bhh}[1]{\ensuremath{\mathcal{B}(\mathcal{H}_{#1})}}

\newcommand{\bhstar}{\ensuremath{\mathcal{B_{*}(H)}}}
\newcommand{\bhhstar}[1]{\ensuremath{\mathcal{B}_{*}(\mathcal{H}_{#1})}}

\newcommand{\kk}{\ensuremath{\mathcal{K}}}
\newcommand{\bkk}{\ensuremath{\mathcal{B(K)}}}

\newcommand{\A}{\ensuremath{\mathcal{A}}}
\newcommand{\B}{\ensuremath{\mathcal{B}}}
\newcommand{\M}{\ensuremath{\mathcal{M}}}

\newcommand{\Bstar}{\ensuremath{\mathcal{B_{*}}}}
\newcommand{\Mstar}{\ensuremath{\mathcal{M_{*}}}}

\begin{document}

\title{Quantum Channels with Memory}

\author{Dennis Kretschmann}
\email{d.kretschmann@tu-bs.de}
\author{Reinhard F. Werner}
\email{r.werner@tu-bs.de}
\affiliation{Institut f\"ur
Mathematische Physik, Technische
    Universit\"at Braunschweig, Mendelssohnstr.~3, D-38106 Braunschweig, Germany}
\date{9~May~2005}

\begin{abstract}
    We present a general model for quantum channels with memory,
    and show that it is sufficiently general to encompass all {\em
    causal} automata: any quantum process in which outputs up to
    some time $t$ do not depend on inputs at times $t' >t$ can be
    decomposed into a concatenated memory channel. We then examine
    and present different physical setups in which channels with
    memory may be operated for the transfer of (private) classical
    and quantum information. These include setups in which either
    the receiver or a malicious third party have control of the
    initializing memory. We introduce classical and quantum
    channel capacities for these settings, and give several
    examples to show that they may or may not coincide. Entropic
    upper bounds on the various channel capacities are given.
    For {\em forgetful} quantum channels, in which the effect of
    the initializing memory dies out as time increases, coding
    theorems are presented to show that these bounds may be
    saturated. Forgetful quantum channels are shown to be open and
    dense in the set of quantum memory channels.
\end{abstract}



\maketitle \tableofcontents

%%%%%%%%%%%%%%%%%%%%%%%%%%%%%%%%%%%%%%%%%%%%%%%%%%%%%%%%%%%%%%%%%%%%%%%%%%%%%%%%%%
%%%%%%%%%%%%%%%%%%%%%%%%%%%%%%%%%%%%%%%%%%%%%%%%%%%%%%%%%%%%%%%%%%%%%%%%%%%%%%%%%%
%%%%%%%%%%%%%%%%%%%%%%%%%%%%%%%%%%%%%%%%%%%%%%%%%%%%%%%%%%%%%%%%%%%%%%%%%%%%%%%%%%
%%%%%%%%%%%%%%%%%%%%%%%%%%%%%%%%%%%%%%%%%%%%%%%%%%%%%%%%%%%%%%%%%%%%%%%%%%%%%%%%%%

\section{Introduction}
    \label{sec:intro}

Any processing of quantum information, be it storage or transfer,
can be represented as a quantum channel: a completely positive and
trace-preserving map $S$ that transforms states (density matrices)
on the sender's end of the channel into states on the receiver's
end. Until now most of the work on quantum channels has
concentrated on {\em memoryless} channels, which are characterized
by the requirement that successive channel inputs are acted on
independently. Mathematically, this means that messages of $n$
symbols are processed by the tensor product channel $S^{\otimes
n}$.

However, in many real-world applications the assumption of having
uncorrelated noise channels cannot be justified, and {\em memory
effects} need to be taken into account. It thus seems desirable to
extend the theory of quantum channels to encompass memory effects,
and to create a common framework in which experiments with both
correlated and uncorrelated noise can be naturally described. In
fact, such a framework is already necessary for estimates on
almost memoryless channels, for instance when assessing whether a
particular system can arguably be modelled as a memoryless
channel. In the present paper such a unified framework will be
presented, and it will be shown how this model can be applied to
the description of different information processing tasks, such as
(private) classical and quantum information transfer.

%%%%%%%%%%%%%%%%%%%%%%%%%%%%%%%%%%%%%%%%%%%%%%%%%%%%%%%%%%%%%%%%%%%%%%%%%%%%%%%%%%
%%%%%%%%%%%%%%%%%%%%%%%%%%%%%%%%%%%%%%%%%%%%%%%%%%%%%%%%%%%%%%%%%%%%%%%%%%%%%%%%%%

\subsection{Outline and Overview}
    \label{sec:outline}

\begin{figure}
    \begin{center}
        \psfrag{t}[cc][cc]{time}
        \psfrag{S}[cc][cc]{\bf S}
        \psfrag{A}{$\mathcal{A}$}
        \psfrag{B}{$\mathcal{B}$}
        \psfrag{M}{$\mathcal{M}$}
        \includegraphics[width=\columnwidth]{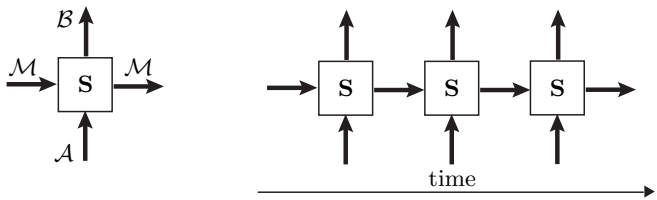}
        \caption{\label{fig:memory} {\em Left:} A quantum memory
        channel with input register $\mathcal{A}$, output register
        $\mathcal{B}$, and memory system $\mathcal{M.}$ --- {\em
        Right:} A threefold concatenation $S_3$ of memory channels,
        with time running from left to right, and coded information
        running from bottom to top.}
    \end{center}
\end{figure}

In our contribution we present a general model for quantum
channels with memory. In addition to Alice's input register
$\mathcal{A}$ and Bob's output register $\mathcal{B}$, such a
channel has an additional memory input and an additional memory
output, denoted by $\mathcal{M}$ (cf. Fig.~\ref{fig:memory},
left). Long messages with $n$ signal states will then be processed
by subsequent application of these memory channels, resulting in
the concatenated channel $S_{n}$ depicted in Fig.~\ref{fig:memory}
(right). This picture will be turned into a rigorous definition in
Section~\ref{sec:constructive}, after the mathematical framework
will have been introduced in Section~\ref{sec:notations}.

In such a setup, the memory system is passed on from one
application of the channel to the next, and introduces (quantum or
classical) correlations between consecutive signal states. If no
memory system is present, the concatenated channel will simply be
a product channel, bringing us back to the memoryless realm in
which consecutive signal states are acted on independently.\\

This model marks a {\em constructive approach} to quantum channels
with memory. It is certainly the appropriate framework when the
physical realization of the memory $\mathcal{M}$ is known.
However, in many applications of information theory only the
input-output behavior of a channel is of interest. From this point
of view the memory would be part of the internal workings of the
channel, and would not be made part of the description. We call
this way of describing channels the {\em axiomatic approach}: It
takes a channel as a transformation turning infinite strings of
input systems to infinite strings of outputs, with only two basic
assumptions: translation invariance and the condition of {\em
causality}. Outputs up to some time $t$ do not depend on inputs at
times $t'> t$. In the classical theory, such channels are
sometimes called {\em non-anticipatory}. It is clear from
Fig.~\ref{fig:memory} that a channel with memory automatically
satisfies this causality condition.

\begin{figure}
    \begin{center}
        \psfrag{t}[cc][cc]{time}
        \psfrag{S}[cc][cc]{\bf S}
        \psfrag{R}[cc][cc]{\bf R}
        \psfrag{T}[cc][cc]{\bf T}
        \psfrag{o}[cc][cc]{$\mathbf{1}$}
        \psfrag{=}[cc][cc]{\large $=$}
        \includegraphics*[width=\columnwidth,bb = 20 290 574 455]{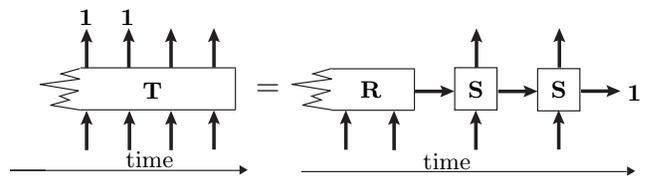}
        \caption{\label{fig:structure} By the {\em Structure Theorem},
        a causal automaton $T$ can be decomposed into a chain of
        concatenated memory channels $S$ plus some input
        initializer $R$. Evaluation with the identity operator $\mathbf{1}$
        means that the corresponding output is ignored.}
    \end{center}
\end{figure}

Taking a causal channel and representing it as a channel with
memory amounts to reconstructing a model of the channel and its
internal memory states and dynamics. This is a highly non-trivial
task, even in the classical case. However, a formal reconstruction
can always be given. This is what we call the {\em Structure
Theorem} for causal channels, and is illustrated in
Fig.~\ref{fig:structure}. A rigorous version will be given as
Th.~\ref{theo:structure} in Section~\ref{sec:causal}. In general,
it produces not only the channel step operator $S$, but also a map
$R$ defining the influence of input states in the remote past on
the memory. Intuitively, however, such a map is often not needed,
because memory effects decrease in time. A similar condition is
needed for passing from the constructive approach of channels with
memory to causal input-output channels: Since the constructive
approach allows one to choose the initial memory state, output
states in general depend on this choice, and in general this
influence will depend on the time after initialization. So in
order to get a time translation invariant channel without such
dependence, the channel $S$ must lose the initialization
information. We call $S$ {\em forgetful} if outputs at a large
time $t$ depend only weakly on the memory initialization at time
zero, in a sense made precise in Section~\ref{sec:forgetful}. For
forgetful channels, memory effects will be shown to decrease even
exponentially.

Not every channel is forgetful. The prime counterexample is a
channel with a {\em global classical switch} discussed in
Section~\ref{sec:examples}. The memory in this case is a classical
bit, left unchanged by $S$, but determining which of two
memoryless channels $S_0,S_1$ is applied to the input at each
time. However, we will show in Section~\ref{sec:forgetful} that
generic memory channels are in fact forgetful, in the sense that
every non-forgetful quantum channel can be approximated by a
forgetful channel to arbitrary degree of accuracy. In addition,
for every forgetful quantum channel we may find a finite-size
neighborhood in which all channels are likewise forgetful. In
mathematical terms, forgetful quantum channels are both {\em open}
and {\em dense} in the set of quantum memory channels.\\

\begin{figure}
    \begin{center}
        \psfrag{E}[cc][cc]{Encoding}
        \psfrag{S}[cc][cc]{\bf S}
        \psfrag{D}[cc][cc]{Decoding}
        \psfrag{=}[cc][cc]{\large $\approx$}
        \includegraphics*[width=1.3\columnwidth,bb=10 256 654 504]{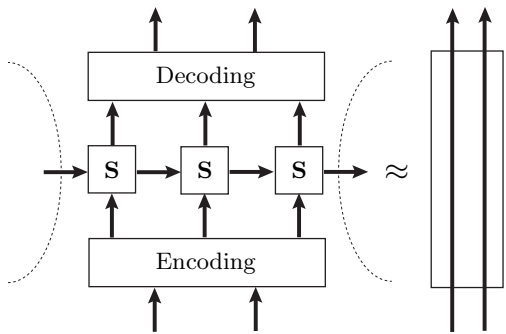}
        \caption{\label{fig:concatenate} Two signal states are encoded
        into three input registers, sent through the concatenated
        memory channel, and then decoded into two output states.
        If the overall channel is (in some sense to be specified
        in Section~\ref{sec:capacity}) close
        to the ideal channel on two inputs, the transmission rate
        of the above scheme is $\frac{2}{3}$. Capacity is the
        largest such rate, in the limit of long messages and
        optimal encoding and decoding. In the above setup, the
        initial memory input can be thought of as being controlled
        by either the sender or a malicious third party.
        Similarly, the receiver may or may not be able to read out
        the final memory state.}
    \end{center}
\end{figure}

For quantum channels with memory, capacity can be defined along
the lines familiar from the memoryless setting \cite{Key02,KW04},
both for the transmission of classical and quantum information.
Channel capacity expresses quantitatively how well a given channel
$S$ can simulate a noiseless qubit (or bit) channel: roughly
speaking, it is the maximal number of ideal qubit (resp. bit)
transmissions per use of the channel, taken in the limit of long
messages and using encoding and decoding schemes asymptotically
eliminating all errors. The concept is illustrated in
Fig.~\ref{fig:concatenate}.

However, when trying to send information through a concatenated
memory channel, unlike in the memoryless case we also have to
specify how to handle the initial and final memory state. In
particular, we may distinguish between setups in which Alice can
access the initial memory input state and may use it for the
encoding procedure, and setups in which a malicious third party
(Eve, say) controls the initial memory input, and by her choice of
the input state will try to prevent Alice and Bob from
communicating over the channel. Likewise, we may consider setups
in which either Bob or Eve control the final memory output. These
distinctions will be made precise in Section~\ref{sec:capacity}.
They lead to slight variations in the notion of capacity, and in
Section~\ref{sec:examples} we will present several examples to
show that the resulting capacities may or may not coincide. In
particular, for channels with only one Kraus operator, all these
capacities are the same, and equal the capacity of the ideal
channel (cf. Section~\ref{sec:pure}).\\

The various capacities can be bounded from above both in terms of
the capacity of memoryless channels and in terms of entropic
expressions. Some of these bounds will be presented. In
particular, the standard mutual information and coherent
information bounds familiar from the memoryless setting easily
extend to memory channels (cf. Section~\ref{sec:bounds}).\\

Forgetful channels are, in a sense to be specified in
Section~\ref{sec:forgetful}, close to memoryless channels. As
such, they play a central role not only as the bridge between the
axiomatic and the constructive approach to quantum memory channels
and as generic examples for quantum memory channels, but also
connect them to the memoryless realm. In Section~\ref{sec:coding},
we will explain how the standard random coding techniques familiar
from the memoryless setting can be modified to saturate the
entropic upper bounds on the channel capacity for forgetful
channels, leading to coding theorems for (private) classical and
quantum information transfer for this very important class of memory channels.\\

We conclude with a Summary and Outlook. An Appendix contains some
mathematical background relevant to the description of
infinite-dimensional quantum systems, insofar as it is essential
to the understanding of the Structure Theorem.

%%%%%%%%%%%%%%%%%%%%%%%%%%%%%%%%%%%%%%%%%%%%%%%%%%%%%%%%%%%%%%%%%%%%%%%%%%%%%%%%%%
%%%%%%%%%%%%%%%%%%%%%%%%%%%%%%%%%%%%%%%%%%%%%%%%%%%%%%%%%%%%%%%%%%%%%%%%%%%%%%%%%%

\subsection{Model Systems and Related Work}
    \label{sec:related}

\begin{figure}
    \begin{center}
        \psfrag{A}[cc][cc]{$\mathcal{A}$}
        \psfrag{B}[cc][cc]{$\mathcal{B}$}
        \includegraphics[width=0.7\columnwidth]{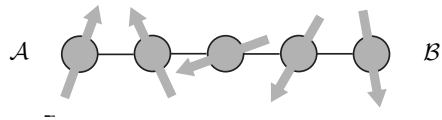}
        \caption{\label{fig:spinchain} An unmodulated spin chain as a quantum
        channel with memory: {\underline A}lice places the input
        signal on the first spin of the chain and lets it
        propagate to {\underline B}ob, who controls the spin at
        the opposite end of the chain.}
    \end{center}
\end{figure}

Quantum channels which naturally acquire a memory are abundant in
all branches of quantum information processing:

Recently, an unmodulated spin chain has been proposed as a model
for short distance quantum communication
\cite{Bos03,BB04,CDE+04,CDD+04}. In such a scheme, the state to be
communicated over the channel is placed on one of the spins of the
chain, propagates for a specific amount of time, and is then
received at a distant spin of the chain (cf. Fig.
\ref{fig:spinchain}). When viewed as a model for quantum
communication, it is generally assumed that a reset of the spin
chain occurs after each signal \cite{GF04}, for example by
applying an external magnetic field, resulting in a memoryless
channel. However, a continuous operation without reset may lead to
higher transmission rates, and corresponds to a quantum channel
with memory.

\begin{figure}
    \begin{center}
        \includegraphics*[width=0.7\columnwidth]{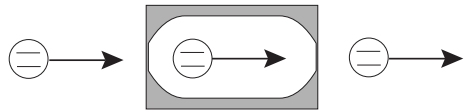}
        \caption{\label{fig:maser} In a {\em micromaser}, a stream
        of two-level atoms is injected into a high-quality
        superconducting cavity. The field modes introduce
        correlations between consecutive atoms.}
    \end{center}
\end{figure}

Another model of a quantum channel with memory is the so-called
{\em one-atom maser} or {\em micromaser} \cite{MWM85,VBW+00}. In
such a device, excited atoms interact with the photon field inside
a high-quality optical cavity, as depicted in Fig.
\ref{fig:maser}. If the photons inside the cavity have
sufficiently long lifetime, atoms entering the cavity will feel
the effect of the preceding atoms, introducing correlations
between consecutive signal
states.\\

Apparently, the first model of a quantum channel with memory was
introduced by Macchiavello et al. in 2001: they gave an example of
a qubit channel with Markovian correlated noise \cite{MP02,MPV03}
in which entangled input states may increase the transmission rate
for classical information. These results have recently been
extended to some bosonic Gaussian channels \cite{CCM+04,RSG+05}.
Such an effect has been demonstrated experimentally for optical
fiber channels with fluctuating birefringence, in which
consecutive light pulses undergo strongly correlated polarization
transformation \cite{BDB04,BDW+04}. (Whether such examples exist
in the {\em memoryless} setting is still an open question, and
presently considered one of the most eminent open problems of
quantum information theory, with wide implications for other
problems in the field \cite{Sho03,Pom03}.)

Subsequently, the study of quantum channels with memory has
largely been confined to channels with Markovian correlated noise
(cf. \cite{Ham02,BM03} and references therein). A Lindbladian
approach to memory channels has been taken by Daffer et al.
\cite{DWM03,DWC+04}. Upper bounds on the classical capacity for a
more general class of channels have been given recently by Bowen
et al. \cite{BDM03}.

All the memory channels discussed in this Section are causal
quantum channels, and thus the Structure Theorem applies. A
completely different approach has been taken by Hayashi and
Nagaoka \cite{HN03}, who refrain from imposing any structural
assumption on the quantum channels they consider, and apply the
information-spectrum method to obtain a coding theorem for the
classical product state capacity, following work by Verd\'{u} and
Han \cite{VH94} on classical channels with memory.

We refer to Verd\'{u}'s overview paper \cite{Ver98} and the
Gray-Davisson collection \cite{GD77} for more information on
memory channels in the purely classical setting.

%%%%%%%%%%%%%%%%%%%%%%%%%%%%%%%%%%%%%%%%%%%%%%%%%%%%%%%%%%%%%%%%%%%%%%%%%%%%%%%%%%
%%%%%%%%%%%%%%%%%%%%%%%%%%%%%%%%%%%%%%%%%%%%%%%%%%%%%%%%%%%%%%%%%%%%%%%%%%%%%%%%%%
%%%%%%%%%%%%%%%%%%%%%%%%%%%%%%%%%%%%%%%%%%%%%%%%%%%%%%%%%%%%%%%%%%%%%%%%%%%%%%%%%%
%%%%%%%%%%%%%%%%%%%%%%%%%%%%%%%%%%%%%%%%%%%%%%%%%%%%%%%%%%%%%%%%%%%%%%%%%%%%%%%%%%

\section{Language and Notations}
    \label{sec:notations}

\subsection{States, Channels, and Observables}
    \label{sec:channel}

According to the rules of quantum mechanics, every quantum system
is associated with a Hilbert space $\mathcal{H}$, which for the
purpose of this paper can mostly (but not always, see the
discussion in Section~\ref{sec:picture}) be taken as finite
dimensional. The observables of the system are given by bounded
linear operators on the Hilbert space $\mathcal{H}$, written
$\bh$. The physical states associated with the system are density
operators $\varrho \in \bhstar$, where the latter denotes the
space of trace class operators on $\hh$.

A {\em quantum channel} $S$ which transforms input systems
described by a Hilbert space $\hh_{1}$ into output systems
described by a (possibly different) Hilbert space $\hh_{2}$ is
represented mathematically by a completely positive unital map $S
\mathpunct: \bhh{2} \rightarrow \bhh{1}$. By unitality we mean
that $S(\1_{\hh_2}) = \1_{\hh_1}$, with the identity operator
$\1_{\hh_i} \in \bhh{i}$. Each channel $S$ can be written in the
so-called {\em Kraus form} \cite{Kra83}
\begin{equation}
    \label{eqn:kraus}
        S(X) = \sum_{i=1}^{n} s_{i}^{*} X s_{i}^{}
\end{equation}
with a number of $n \leq \dim(\hh_1) \dim(\hh_2)$ {\em Kraus
operators} $s_i \mathpunct: \hh_1 \rightarrow \hh_2$.

The physical interpretation of the quantum channel $S$ is the
following: when the system is initially in the state $\varrho \in
\bhhstar{1}$, the expectation value of the measurement of the
observable $X \in \bhh{2}$ at the output side of the channel is
given in terms of $S$ by $\tr{\varrho \, S(X)}$.

Alternatively, and perhaps more intuitively, we can look at the
dynamics of the states and introduce the dual map $S_{*}
\mathpunct : \bhhstar{1} \rightarrow \bhhstar{2}$ by means of the
duality relation
\begin{equation}
    \label{eqn:dual}
        \tr{S_{*}(\varrho) \, X} = \tr{\varrho \, S(X)}.
\end{equation}
$S_{*}$ is a completely positive and trace-preserving map and
represents the channel in {\em Schr\"odinger picture}, while $S$
provides the {\em Heisenberg picture} representation (cf. Davies'
textbook \cite{Dav76} and Keyl's survey article \cite{Key02} for a
more extensive discussion of observables, states, and channels).

%%%%%%%%%%%%%%%%%%%%%%%%%%%%%%%%%%%%%%%%%%%%%%%%%%%%%%%%%%%%%%%%%%%%%%%%%%%%%%%%%%
%%%%%%%%%%%%%%%%%%%%%%%%%%%%%%%%%%%%%%%%%%%%%%%%%%%%%%%%%%%%%%%%%%%%%%%%%%%%%%%%%%

\subsection{Heisenberg vs. Schr\"odinger}
    \label{sec:picture}

For the finite dimensional systems we will consider in
Section~\ref{sec:memory}, Schr\"odinger picture and Heisenberg
picture are completely equivalent descriptions of quantum
processes by means of the duality relation Eq.~(\ref{eqn:dual}).
However, in the axiomatic characterization of quantum channels, as
presented in Section~\ref{sec:causal}, we will have to deal with
infinite-dimensional systems, for which Heisenberg picture is the
mandatory language. Thus, for consistency we work in Heisenberg
picture throughout, emphasizing that for finite-dimensional
systems conversion to Schr\"odinger picture is always immediate
from Eq.~(\ref{eqn:dual}). Some mathematical background on the
description of infinite-dimensional systems, insofar as it is
essential to the understanding of the present paper, is relegated
to the Appendix. Most notably, this includes quasi-local algebras
and Stinespring's dilation theorem.

%%%%%%%%%%%%%%%%%%%%%%%%%%%%%%%%%%%%%%%%%%%%%%%%%%%%%%%%%%%%%%%%%%%%%%%%%%%%%%%%%%
%%%%%%%%%%%%%%%%%%%%%%%%%%%%%%%%%%%%%%%%%%%%%%%%%%%%%%%%%%%%%%%%%%%%%%%%%%%%%%%%%%

\subsection{Distance between Quantum Channels}
    \label{sec:distance}

From the informal discussion in Section~\ref{sec:outline} it is
clear that the definition of channel capacity requires the
comparison of the quantum channel after the encoding and decoding
process with an ideal channel. As a measure of the distance
between two channels we favor the {\em norm of complete
boundedness}, (or {\em cb-norm}, for short) \cite{Pau02}, denoted
by $\cb{\cdot}$. For two channels $T$ and $S$, the distance
$\frac12\cb{T-S}$ can be defined as the largest difference between
the overall probabilities in two statistical quantum experiments
differing only by exchanging one use of $S$ by one use of $T$.
These experiments may involve entangling the systems on which the
channels act with arbitrary further systems. Equivalently, we may
set $\cb T=\sup_n\norm{T\otimes\id_n}$, where $\norm{\cdot}$
denotes the norm of linear operators between the Banach spaces
$\bhh{i}$ (cf. Appendix), and $\id_n$ denotes the identity map
(ideal channel) on the $n\times n$ matrices.

Among the properties which make the  cb-norm well-suited for
capacity estimates are norm multiplicativity,  $\cb{T_1 \otimes
T_2} \, = \, \cb{T_1} \, \cb{T_2}$, and unitality, $\cb{T} \, = \,
1$ for any channel $T$. The equivalence with other error criteria
such as {\em minimum fidelity} and {\em entanglement fidelity} is
discussed extensively in \cite{KW04}.

When working in the Schr\"odinger picture representation, the
so-called {\em trace norm} $\tracenorm{\varrho} := {\rm tr } \,
\sqrt{\varrho^* \varrho}$ is frequently used to evaluate the
distance between two quantum states. Again we refer to \cite{KW04}
for the equivalence with other distance measures.\\

Note that throughout this work we use base two logarithms, and we
write $\ld x := \log_2 x$.

%%%%%%%%%%%%%%%%%%%%%%%%%%%%%%%%%%%%%%%%%%%%%%%%%%%%%%%%%%%%%%%%%%%%%%%%%%%%%%%%%%
%%%%%%%%%%%%%%%%%%%%%%%%%%%%%%%%%%%%%%%%%%%%%%%%%%%%%%%%%%%%%%%%%%%%%%%%%%%%%%%%%%
%%%%%%%%%%%%%%%%%%%%%%%%%%%%%%%%%%%%%%%%%%%%%%%%%%%%%%%%%%%%%%%%%%%%%%%%%%%%%%%%%%
%%%%%%%%%%%%%%%%%%%%%%%%%%%%%%%%%%%%%%%%%%%%%%%%%%%%%%%%%%%%%%%%%%%%%%%%%%%%%%%%%%

\section{Channels with Memory}
    \label{sec:memory}

\subsection{The Constructive Approach}
    \label{sec:constructive}

A relatively simple (yet surprisingly general, see below) model to
describe channels with correlated noise consists of a quantum
channel which, in addition to Alice's input register system
$\hh_A$ and Bob's output register system $\hh_B$ has an additional
memory input $\hh_M$ and an additional memory output $\hh_{M'}$.
(Since the smaller of the two Hilbert space $\hh_M$, $\hh_{M'}$
can always be thought of as being embedded in the larger one, in
the following we will assume without loss that $\hh_M =
\hh_{M'}$.) Mathematically, a {\em quantum channel with memory}
(or, for short, {\em memory channel}) is represented (in
Heisenberg picture) as a completely positive and unital map $S
\mathpunct : \bhh{B} \otimes \bhh{M} \rightarrow \bhh{M} \otimes
\bhh{A}$. Often we will abbreviate $\bhh{A}$ to $\A$, and
similarly for $\bhh{B}$ and $\bhh{M}$. Long messages with $n \in
\N$ signal states will then be processed by subsequent application
of memory channels, resulting in the concatenated channel $S_n
\mathpunct : \B^{\otimes n} \otimes \M \rightarrow \M \otimes
\A^{\otimes n}$ given as follows (see Fig. \ref{fig:memory}):
\begin{equation}
    \label{eqn:constructive01}
        S_n =  \big ( S \otimes \id_{\A}^{\otimes n-1} \big ) \circ ...
        \circ \big ( \id_{\B}^{\otimes n-2} \otimes S \otimes
        \id_{\A}^{} \big ) \circ \big ( \id_{\B}^{n-1} \otimes S
        \big),
\end{equation}
where $\id$ denotes the identity operation ({\em ideal} or {\em
noiseless} channel): $\id(X) = X \; \forall \; X$.

The Schr\"odinger picture equivalent of this model was introduced
by Bowen and Mancini in \cite{BM03} and has been shown to
encompass channels with Markovian correlated noise discussed
previously in \cite{MP02,MPV03,DWM03,BDB04}. As advertised in the
Introduction, in Section \ref{sec:causal} we will show that this
model is sufficiently general to describe all causal quantum
channel, which was left as an open problem in \cite{BM03}.
However, before we prove the Structure Theorem we will extend the
notion of channel capacity from the memoryless setting to channels
with memory, and we will present several different setups in which
these channels may be operated for the transmission of both
classical and quantum information.

%%%%%%%%%%%%%%%%%%%%%%%%%%%%%%%%%%%%%%%%%%%%%%%%%%%%%%%%%%%%%%%%%%%%%%%%%%%%%%%%%%
%%%%%%%%%%%%%%%%%%%%%%%%%%%%%%%%%%%%%%%%%%%%%%%%%%%%%%%%%%%%%%%%%%%%%%%%%%%%%%%%%%

\subsection{Channel Capacity}
    \label{sec:capacity}

As explained in Section~\ref{sec:outline}, the standard definition
of capacity applies also to quantum channels with memory. However,
as illustrated in Fig.~\ref{fig:concatenate} we have to specify
how to handle the initial and final memory states. In particular,
we need to distinguish between setups in which Alice has control
over the initial memory input state and may use it for the
encoding procedure, and setups in which a malicious third party
(Eve, say) controls the initial memory input, and by her choice of
the input state $\mu \in \bhhstar{M}$ will try to prevent Alice
and Bob from communicating over the channel. Likewise, we may
consider setups in which the final memory states is either ignored
or accessible to Bob, and can thus be employed in the decoding
process.

In the definition of channel capacity presented below, these four
different scenarios are distinguished by a different range and
domain of the encoding and decoding map, respectively, and give
rise to four different channel capacities for both classical and
quantum information transmission.

%%%%%%%%%%%%%%%%%%%%%%%%%%%%%%%%%%%%%%%%%%%%%%%%%%%%%%%%%%%%%%%%%%%%%%%%%%%%%%%%%%
\begin{define}
    \label{def:cap}
        Let $\hh_A$, $\hh_B$, and $\hh_M$ be Hilbert spaces.
        A positive number $R$ is called an {\em achievable rate} for
        the quantum memory channel $S \mathpunct : \bhh{B} \otimes
        \bhh{M} \rightarrow \bhh{M} \otimes \bhh{A}$ iff for any
        pair of integer sequences $(n_{\nu})_{\nu\in\N}$ and
        $(m_{\nu})_{\nu\in\N}$ with $\lim_{\nu\to\infty} n_\nu \,
        = \, \infty$ and $\varlimsup_{\nu\to\infty}
        \frac{m_\nu}{n_\nu} \, \leq \, R$ we have
        \begin{equation}
            \label{eqn:cap01}
            \lim_{\nu\to\infty} \, \Delta(n_\nu,m_\nu) \, = \, 0,
        \end{equation}
        where we set
        \begin{equation}
            \label{eqn:cap02}
            \Delta(n_\nu,m_\nu) \, := \, \inf_{E,D} \, \cb{E \, S_{n_\nu}
            D \, - \, \id_{\C^2}^{\otimes m_\nu} \,},
        \end{equation}
        the infimum taken over all encoding channels $E$ and decoding channels
        $D$ with suitable domain and range.\\ The {\em quantum
        channel capacity} $Q(S)$ of the memory channel $S$ is
        defined to be the supremum of all achievable rates.\\ In the
        different setups described above, the domain of the encoding channels $E$
        may or may not include the initial memory algebra
        $\bhh{M}$, and the range of the decoding channels $D$ may
        or may not contain the final memory algebra $\bhh{M}$,
        resulting in four different quantum capacities $Q_{AB}(S)$,
        $Q_{AE}(S)$, $Q_{EB,\mu}(S)$, and $Q_{EE,\mu}(S)$, where the first
        index stands for the party ({\underline A}lice, {\underline B}ob, or
        {\underline E}ve) who controls the initial memory state, the
        second index stands for the party who has access to the
        final memory state, and $\mu \in \bhhstar{M}$ stands for
        Eve's choice of the initial memory state, if applicable.
\end{define}
%%%%%%%%%%%%%%%%%%%%%%%%%%%%%%%%%%%%%%%%%%%%%%%%%%%%%%%%%%%%%%%%%%%%%%%%%%%%%%%%%%
\begin{remark}
    \label{remark:classical}
        {\em The capacity of a quantum memory channel $S$ for the transmission
        of {\em classical} information can be defined along the same
        lines, restricting encoding channels to {\em preparations} and
        decoding channels to {\em measurements} \cite{Wer01}, and
        replacing the ideal qubit channel $\id_{\C^2}$ by the ideal bit channel in
        Eq.~(\ref{eqn:cap02}). The respective capacities are denoted by
        $C_{AB}(S)$, $C_{AE}(S)$, $C_{EB,\mu}(S)$, and $C_{EE,\mu}(S)$, and
        are no smaller than their quantum counterparts.}
\end{remark}
%%%%%%%%%%%%%%%%%%%%%%%%%%%%%%%%%%%%%%%%%%%%%%%%%%%%%%%%%%%%%%%%%%%%%%%%%%%%%%%%%%
\begin{remark}
    \label{remark:star}
        {\em In the sections to follow, we will write $Q_{*}(S)$ and
        $C_{*}(S)$ whenever a certain statement holds for all the
        four channel capacities introduced in Def.~\ref{def:cap},
        regardless of Eve's choice of the initial memory state.}
\end{remark}
%%%%%%%%%%%%%%%%%%%%%%%%%%%%%%%%%%%%%%%%%%%%%%%%%%%%%%%%%%%%%%%%%%%%%%%%%%%%%%%%%%
\begin{remark}
    \label{remark:relations}
        {\em It is obvious from the definition that for every memory channel
        $S$ the capacities introduced in Def.~\ref{def:cap} satisfy the
        following chain of inequalities:
        \begin{equation}
            \label{eqn:relations}
                Q_{EE,\mu}(S) \leq \left \{ Q_{AE}(S), Q_{EB,\mu} \right \} \leq Q_{AB}(S)
        \end{equation}
        for all $\mu \in \bhhstar{M}$, and accordingly for the classical
        capacities $C_{EE,\mu}(S)$ etc.}
\end{remark}
%%%%%%%%%%%%%%%%%%%%%%%%%%%%%%%%%%%%%%%%%%%%%%%%%%%%%%%%%%%%%%%%%%%%%%%%%%%%%%%%%%
\begin{remark}
    \label{remark:equivalent}
        {\em Note that there are several equivalent definitions of
        channel capacity. In particular, it is
        sufficient to find one pair of integer sequences $(n_{\nu})_{\nu\in\N}$ and
        $(m_{\nu})_{\nu\in\N}$ such that $\varlimsup_{\nu\to\infty}
        \frac{m_\nu}{n_\nu} = R$ and $\lim_{\nu\to\infty} \, \Delta(n_\nu,m_\nu) =
        0$, provided the diverging sequence $(n_{\nu})_{\nu\in\N}$
        is {\em subexponential}, i.\,e., $\lim_{\nu\to\infty}
        \frac{n_{\nu + 1}}{n_\nu} = 1.$\\ In addition, the cb-norm
        in Eq.~(\ref{eqn:cap02}) can be replaced by other
        distance measures such as {\em minimum
        fidelity} or {\em entanglement fidelity}. See \cite{KW04} for a detailed
        discussion of these matters.}
\end{remark}

%%%%%%%%%%%%%%%%%%%%%%%%%%%%%%%%%%%%%%%%%%%%%%%%%%%%%%%%%%%%%%%%%%%%%%%%%%%%%%%%%%
%%%%%%%%%%%%%%%%%%%%%%%%%%%%%%%%%%%%%%%%%%%%%%%%%%%%%%%%%%%%%%%%%%%%%%%%%%%%%%%%%%

\subsection{Examples}
    \label{sec:examples}

In the following, in order to illustrate the concepts introduced
above we will present several examples of quantum memory channels.
These examples will also serve to show that the different
capacities introduced in Def.~\ref{def:cap} may or may not
coincide, thereby justifying our defining more than one capacity.

A simple model channel for which all the capacities introduced
above coincide is the {\em Shift Channel} $S^s$. In principle,
this is just a noiseless channel, but it interchanges memory and
input register: $S^{s}(b \otimes m) = b \otimes m$ (Note that in
the tensor representation that we have chosen, the identity
channel $\id$ comes with the inherent flip, i.\,e., $\id (b
\otimes m) = m \otimes b$.) Thus, in an $n$-fold concatenation of
Shift Channels, the signals that Alice sends through the channel
will be received by Bob undistorted one time-step later. In the
capacity limit of long messages, as $n\to\infty$, the initial
qubit that Bob may lose if Eve controls the initial memory state,
and the final qubit that he may lose if he cannot access the final
memory state both have a negligible impact on the transmission
rate, and therefore $Q_{EE,\mu} (S^{s}) = \lim_{n\to\infty}
\frac{n-2}{n} \ld d = \ld d \; \forall \; \mu \in \bhhstar{M}$,
with $d:= \dim \hh_A = \dim \hh_B = \dim \hh_M$. Therefore, by
Eq.~(\ref{eqn:relations}) and Remark~\ref{remark:classical} all
the above capacities equal $\ld d$.  Further examples for channels
in which the worst-case capacity and the best-case capacity are
both maximal will be presented in Section~\ref{sec:pure}.

An example of a memory channel in which the control over the
initializing memory state can have a decisive influence on the
channel performance is the channel with a {\em global classical
switch}: Suppose that the memory algebra is a classical $d$-level
system of diagonal $d \times d$ matrices, and that we are given a
collection $\{T_i \}_{i = 1}^{d}$ of $d$ quantum memoryless
channels $T_i \mathpunct : \B \rightarrow \A$. Then a quantum
memory channel $S \mathpunct : \B \otimes \M \rightarrow \M
\otimes \A$ with a global classical switch ($d$ settings) is given
by
\begin{equation}
    \label{eqn:switch}
        S(b \otimes m) = \sum_{i=1}^{d} \bra{i} m \ket{i} \; \kb{i} \otimes T_{i}
        (b).
\end{equation}
In an $n$-fold concatenation of this channel, the channel $T_i$ is
applied in every time step if the initial memory input state was
$\kb{i}$. If Alice initially sends a pre-defined sequence of test
states, Bob may find out what the initial memory setting was and
choose the decoding channel accordingly. Thus, the best case
capacity in this setting will be $\max_{i=1,..d} \left \{ Q(T_i)
\right \}$, and the worst case capacity will be no larger than
$\min_{i=1,..d} \left \{ Q(T_i) \right \}$. These two may clearly
differ.

%%%%%%%%%%%%%%%%%%%%%%%%%%%%%%%%%%%%%%%%%%%%%%%%%%%%%%%%%%%%%%%%%%%%%%%%%%%%%%%%%%
%%%%%%%%%%%%%%%%%%%%%%%%%%%%%%%%%%%%%%%%%%%%%%%%%%%%%%%%%%%%%%%%%%%%%%%%%%%%%%%%%%

\subsection{Pure Channels}
    \label{sec:pure}

Pure memory channels are channels which have only one Kraus
operator in Eq.~(\ref{eqn:kraus}). From the unitality condition,
$S(\1) = \1$, it is then clear that these channels have a Kraus
representation $S(b \otimes m) = V^{*} (b \otimes m) V^{}$ with
isometric $V: \hh_{M} \otimes \hh_{A} \rightarrow \hh_{B} \otimes
\hh_{M}$.

In this section we will show that for pure channels with finite
memory, the various capacities introduced in Def.~\ref{def:cap}
coincide and are maximal, i.\,e., we have the following

%%%%%%%%%%%%%%%%%%%%%%%%%%%%%%%%%%%%%%%%%%%%%%%%%%%%%%%%%%%%%%%%%%%%%%%%%%%%%%%%%%
\begin{theo}
    \label{theo:pure}
        Let $S \mathpunct : \bhh{B} \otimes
        \bhh{M} \rightarrow \bhh{M} \otimes \bhh{A}$ be a pure
        quantum memory channel with finite memory algebra $\bhh{M}$.
        With the convention introduced in
        Remark~\ref{remark:star} we then have:
        \begin{equation}
            \label{eqn:pure01}
                Q_{*}(S) = \min \left \{ \ld \dim \hh_A, \ld \dim \hh_B \right
                \} = C_{*}(S).
        \end{equation}
\end{theo}
%%%%%%%%%%%%%%%%%%%%%%%%%%%%%%%%%%%%%%%%%%%%%%%%%%%%%%%%%%%%%%%%%%%%%%%%%%%%%%%%%%
Our strategy for the proof is to show that for pure channels it is
possible to satisfy the Knill-Laflamme error correction criteria
\cite{KL97}, which imply that perfect signal recovery can be
achieved. This is even more than what is required for capacity
purposes, since the definition of channel capacity, as presented
in Section~\ref{sec:capacity}, only demands that errors vanish
asymptotically, i.\,e., in the limit of long messages
$n\to\infty$.

Since we will have to refer to them repeatedly in the course of
the proof, we start by restating the Knill-Laflamme conditions for
perfect error correction (cf. Th. 10.1 in \cite{NC00}): A
necessary and sufficient condition for a quantum channel $T
\mathpunct : \bhh{2} \rightarrow \bhh{1}$ with Kraus operators
$\left \{t_{i} \right \}_{i=1}^{K}$ to be completely correctable
on a subspace $\kk \subset \hh_1$ is the existence of an
orthonormal basis $\left \{ \ket{\alpha} \right
\}_{\alpha=1}^{\dim \kk}$ of $\kk$ such that
\begin{equation}
    \label{eqn:kl}
        \bra{\alpha} t_{i}^{*} t_{j}^{} \ket{\beta} = \omega_{i,j}
        \; \ip{\alpha}{\beta},
\end{equation}
where the coefficients $\omega_{i,j} \in \C$ are not permitted to
depend on the basis labels $\alpha$, $\beta$. If the orthonormal
basis $\left \{ \ket{\alpha} \right \}_{\alpha} \subset \hh_1$ has
$N$ elements, we say that there exists a quantum code of dimension
$N$.

Coming back to pure channels, we see that in the setup in which
Alice controls the initial memory state and Bob can read out the
final memory state there is only one (isometric) Kraus operator
$V$, and thus it is straightforward to satisfy Eq.~(\ref{eqn:kl})
and achieve rates of up to $\min \left \{ \ld \dim \hh_A, \ld \dim
\hh_B \right \}$.

By Eq.~(\ref{eqn:relations}) and Remark~\ref{remark:classical}, in
order to complete the proof of Th.~\ref{theo:pure} it is therefore
sufficient to show that $Q_{EE,\mu} \geq \min \left \{ \ld \dim
\hh_A, \ld \dim \hh_B \right \} \; \forall \; \mu \in
\bhhstar{M}$. Again we will show that it is possible to satisfy
the error-correction conditions Eq.~(\ref{eqn:kl}). However, in
the worst-case scenario in which Eve chooses an arbitrary input
state $\mu \in \bhhstar{M}$ and Bob has no control over the final
memory output the resulting channel is no longer pure, but can be
given a Kraus representation with no more than $d_{M}^2$ Kraus
operators, where $d_{M} := \dim \hh_M$:

%%%%%%%%%%%%%%%%%%%%%%%%%%%%%%%%%%%%%%%%%%%%%%%%%%%%%%%%%%%%%%%%%%%%%%%%%%%%%%%%%%
\begin{lemma}
    \label{lemma:pure01}
        Let $\hh_A$, $\hh_B$, and $\hh_M$ be finite-dimensional
        Hilbert spaces, and let $d_{M} := \dim \hh_M$. Suppose
        that $S \mathpunct : \bhh{B} \otimes
        \bhh{M} \rightarrow \bhh{M} \otimes \bhh{A}$ is a pure
        quantum channel, i.\,e., $S(b \otimes m) = V^{*} (b \otimes m) V^{}$ for
        isometric $V: \hh_{M} \otimes \hh_{A} \rightarrow \hh_{B} \otimes
        \hh_{M}$. Let $\hat{S}_{\mu} \mathpunct :
        \bhh{B} \rightarrow \bhh{A}$ be the restriction of $S$ to the
        $B$-system, with fixed initial memory state $\mu \in
        \bhhstar{M}$. Then $\hat{S}_{\mu}$ can be given a
        Kraus representation with $d_{M}^2$ Kraus operators.
\end{lemma}
%%%%%%%%%%%%%%%%%%%%%%%%%%%%%%%%%%%%%%%%%%%%%%%%%%%%%%%%%%%%%%%%%%%%%%%%%%%%%%%%%%
{\bf Proof:} Let $\left \{ \ket{\alpha} \right \}_{\alpha =
1}^{d_M}$ be the eigenbasis of $\mu \in \bhhstar{M}$, and suppose
that $\left \{ \ket{i} \right \}_{i = 1}^{d_A}$ and $\left \{
\ket{j'} \right \}_{j' = 1}^{d_B}$ are orthonormal bases for
$\hh_A$ and $\hh_B$, respectively. The isometry $V: \hh_{M}
\otimes \hh_{A} \rightarrow \hh_{B} \otimes \hh_{M}$ can then be
given the representation
\begin{equation}
    \label{eqn:pure02}
        V = \sum_{\alpha, \beta=1}^{d_M} \; V_{\alpha,\beta} \otimes
        \op{\alpha}{\beta}
\end{equation}
with operators $V_{\alpha,\beta} = \sum_{i=1}^{d_A}
\sum_{j'=1}^{d_B} \; \bra{j',\alpha} V \ket{\beta,i} \,
\op{j'}{i}$. From Eq.~(\ref{eqn:pure02}) we see that for arbitrary
$\varrho \in \bhhstar{A}$ and $b \in \bhh{B}$ we have
\begin{equation}
    \label{eqn:pure03}
        \begin{split}
        {\rm tr} \;  (\varrho & \otimes \mu) \; V^{*} (b \otimes \1_M)
        V^{}\\
        & = \sum_{\alpha,\beta,\gamma=1}^{d_M} \; \tr{\varrho \;
        V_{\alpha,\gamma}^{*} \; b \; V_{\alpha,\beta}^{}} \; \bra{\beta}
        \mu \ket{\gamma} \\
        & = \sum_{\alpha,\beta=1}^{d_M} \mu_{\beta} \; \tr{\varrho \;
        V_{\alpha,\beta}^{*} \; b \; V_{\alpha,\beta}^{}} \\
        & = \sum_{\alpha,\beta=1}^{d_M} {\rm tr } \; \varrho \;
        \hat{s}_{\mu,\alpha \beta}^{*} \; b \; \hat{s}_{\mu,\alpha \beta}^{}\\
        & = {\rm tr} \; \varrho \; \hat{S}_{\mu} (b),
        \end{split}
\end{equation}
where we have set $\hat{s}_{\mu,\alpha \beta} :=
\sqrt{\mu_{\beta}} \, V_{\alpha, \beta}$, and $\left \{
\mu_{\beta} \right \}_{\beta = 1}^{d_M}$ are the eigenvalues of
$\mu \in \bhhstar{M}$. Thus, the restricted channel
$\hat{S}_{\mu}$ can be given a representation with $d_{M}^2$ Kraus
operators, as claimed.
$\blacksquare$\\
%%%%%%%%%%%%%%%%%%%%%%%%%%%%%%%%%%%%%%%%%%%%%%%%%%%%%%%%%%%%%%%%%%%%%%%%%%%%%%%%%%

Note that in this representation the number of Kraus operators is
independent of the dimension of both Alice's and Bob's systems
$\hh_A$ and $\hh_B$, and thus the above result holds true also for
the concatenated memory channel $S_n \mathpunct : \bhh{B}^{\otimes
n} \otimes \bhh{M} \rightarrow \bhh{M} \otimes \bhh{A}^{\otimes
n}$, independently of $n\in\N$. Consequently, in the limit
$n\to\infty$ of long messages our setup corresponds to a channel
with large input space interacting with a small environment.
Physical intuition suggests that in such a setup the loss of
information to the environment should be negligible, and it should
be possible to operate the channel like an almost ideal one. This
is the essence of the following

%%%%%%%%%%%%%%%%%%%%%%%%%%%%%%%%%%%%%%%%%%%%%%%%%%%%%%%%%%%%%%%%%%%%%%%%%%%%%%%%%%
\begin{lemma}
    \label{lemma:pure02}
        Let $T \mathpunct : \bh \rightarrow \bh$ be a channel
        with $K$ Kraus operators. Then there exists a quantum code
        of dimension at least $\left \lfloor \frac{\dim \hh}{2^{K^2}} \right
        \rfloor$.
\end{lemma}
%%%%%%%%%%%%%%%%%%%%%%%%%%%%%%%%%%%%%%%%%%%%%%%%%%%%%%%%%%%%%%%%%%%%%%%%%%%%%%%%%%
{\bf Proof:} Let $\left \{ t_i \right \}_{i=1}^{K}$ be a set of
Kraus operators for $T$, and let $\tau_{i,j} := t^{*}_{i}
t^{}_{j}$. In order to find a subspace $\kk \subset \hh$ of high
dimensionality such that the Knill-Laflamme conditions
Eq.~(\ref{eqn:kl}) are satisfied, the following strategy may seem
promising: Choose a state vector $\varphi_1 \in \hh$ arbitrarily,
and then choose
\begin{equation}
    \label{eqn:pure04}
        \varphi_2 \in \kk_1 := \varphi_{1}^{\perp} \; \cap \;
        \bigcap_{i,j=1}^{K} \; (\tau_{i,j} \varphi_1)^{\perp}.
\end{equation}
Iterate this procedure of successive removal of dimensions until
no further state vectors can be found. In every step, at most
$K^2$ dimensions are removed, so this strategy yields a subspace
of dimension $\geq \frac{\dim \hh}{K^2}$. Unfortunately, this
procedure does not guarantee that inner products
$\bra{\varphi_{\alpha}} \tau_{i,j} \ket{\varphi_{\alpha}}$ are
independent of the basis labels, as required by the Knill-Laflamme
conditions Eq.~(\ref{eqn:kl}). However, this can be accomplished
by a carefully balanced pairing of eigenvectors, at the expense of
a smaller code space:

Note that any operator $\tau \in \bh$ can be written as the
weighted sum of two Hermitian operators, $\tau = \frac{1}{2}
\tau_{+} + \frac{i}{2} \tau_{-}$ with $\tau_{+} := \tau^{*} +
\tau^{}$ and $\tau_{-} := i (\tau^{*} - \tau^{})$. Since the
Knill-Laflamme conditions Eq.~(\ref{eqn:kl}) are linear in the
operators $\tau_{i,j}$, we may assume without loss that all
operators $\tau_{i,j}$ are Hermitian. Let $\tau$ be one of these
operators, and let $\left \{ \lambda_{\alpha} \right
\}_{\alpha=1}^{d}$ be the set of its eigenvalues, where $d := \dim
\hh$ and multiple eigenvalues appear according to their
multiplicity. Choose $\omega \in \R$ such that equally many of the
real numbers $\mu_{\alpha} := \lambda_{\alpha} - \omega$ lie on
the positive and on the negative axis. (If necessary, reduce the
dimension of $\hh$ by one.) Now, if $\psi_{\alpha}$ is some
eigenvector of the operator $\tau - \omega \1$ corresponding to
the eigenvalue $\mu_{\alpha} > 0$, and $\psi_{-\alpha}$ is an
eigenvector corresponding to the eigenvalue $\mu_{-\alpha} < 0$,
by setting
\begin{equation}
    \label{eqn:pure05}
        \varphi_{\alpha} := \frac{1}{\sqrt{1- \frac{\mu_{\alpha}}{\mu_{-\alpha}}}}
        \left ( \psi_{\alpha} + \sqrt{\frac{\mu_{\alpha}}{- \mu_{-\alpha}}}
        \psi_{-\alpha} \right ),
\end{equation}
we obtain a Hilbert space $\kk_1 := {\rm
lin}\left\{\varphi_{\alpha} \; \mid \; \alpha = 1,...,\frac{d}{2}
\right \}$ of dimension $\frac{d}{2}$ satisfying the
Knill-Laflamme conditions Eq.~(\ref{eqn:kl}) for the operator
$\tau$, i.\,e.,
\begin{equation}
    \label{eqn:pure06}
        \bra{\varphi_{\alpha}} \tau - \omega \1 \ket{\varphi_{\beta}} = 0
        \; \; \forall \; \; \alpha, \beta = 1,...,\frac{d}{2}.
\end{equation}
Now, choose another operator $\tau' \in \{ \tau_{i,j}
\}_{i,j=1}^{K}$ and repeat the above pairing procedure on the
subspace $\kk_1$, resulting in a subspace $\kk_2 \subset \hh$ of
dimension $\frac{d}{4}$. After $K^2$ steps, the resulting subspace
has dimension at least $\frac{d}{2^{K^2}}$, which is the desired
result. $\blacksquare$\\
%%%%%%%%%%%%%%%%%%%%%%%%%%%%%%%%%%%%%%%%%%%%%%%%%%%%%%%%%%%%%%%%%%%%%%%%%%%%%%%%%%

We can now complete the proof of Theorem~\ref{theo:pure}: Applying
the Knill-Laflamme code described in the proof of
Lemma~\ref{lemma:pure02} to the concatenated memory channel
$\hat{S}_{\mu,n}$ with $d_{M}^2$ Kraus operators, we immediately
see that for all $\mu \in \bhhstar{M}$
\begin{equation}
    \label{eqn:pure07}
        Q_{EE,\mu} \geq \lim_{n\to\infty}\; \frac{1}{n} \; \ld
        \frac{d^n}{2^{d_{M}^{4}}} = \ld d,
\end{equation}
where $d := \min \left \{ \ld \dim \hh_A, \ld \dim \hh_B \right
\}$, as claimed. $\blacksquare$\\
%%%%%%%%%%%%%%%%%%%%%%%%%%%%%%%%%%%%%%%%%%%%%%%%%%%%%%%%%%%%%%%%%%%%%%%%%%%%%%%%%%

After completion of the present work we learned that closely
related results on channels interacting with {\em small}
environments have been obtained independently by G. Bowen and S.
Mancini \cite{BM04}. These authors also show that for such
channels the Knill-Laflamme error correction conditions can be
fulfilled. However, instead of the pairing of eigenvalues
described in the proof of Lemma~\ref{lemma:pure02}, their approach
uses convex sets arguments of Knill et al. \cite{KLV00}, which are
based on a generalization of Radon's theorem \cite{Tve66}. Our
approach seems more straightforward, but this comes at the expense
of a weaker estimate, since the more sophisticated strategy of
Knill et al. yields a code of dimension $\geq
\frac{d}{K^2(K^{2}+1)}$.

%%%%%%%%%%%%%%%%%%%%%%%%%%%%%%%%%%%%%%%%%%%%%%%%%%%%%%%%%%%%%%%%%%%%%%%%%%%%%%%%%%
%%%%%%%%%%%%%%%%%%%%%%%%%%%%%%%%%%%%%%%%%%%%%%%%%%%%%%%%%%%%%%%%%%%%%%%%%%%%%%%%%%
%%%%%%%%%%%%%%%%%%%%%%%%%%%%%%%%%%%%%%%%%%%%%%%%%%%%%%%%%%%%%%%%%%%%%%%%%%%%%%%%%%
%%%%%%%%%%%%%%%%%%%%%%%%%%%%%%%%%%%%%%%%%%%%%%%%%%%%%%%%%%%%%%%%%%%%%%%%%%%%%%%%%%

\section{The Structure of Causal Channels}
    \label{sec:causal}

In the first part of this work we have followed a constructive
approach to quantum channels with memory, in the sense that
quantum channels which process long messages were always thought
of as concatenations of smaller units which process one quantum
signal each. In this section we take the alternative view and
assume that we are a priori given a quantum channel on a long
(possibly infinite) message string. Our interest is then in the
internal structure of such a quantum channel. As advertised in the
Introduction, we will show in Th.~\ref{theo:structure} that under
very general assumptions it can be decomposed into a chain of
quantum memory channels.

This result requires some mathematical background from the theory
of infinite-dimensional quantum systems and channel
representations, most notably quasi-local algebras and the
uniqueness of the minimal Stinespring dilation. The relevant
material is collected in the Appendix.\\

To set the stage, imagine that we have at our disposal a quantum
channel which, at every discrete time step, transforms an input
state on some observable algebra $\A$ into an output state on some
(possibly different) observable algebra $\B$. It is represented
(in Heisenberg picture) by a completely positive and unital map $T
\mathpunct : \B_{\Z} \rightarrow \A_{\Z}$ between the quasi-local
algebras $\A_{\Z}$ and $\B_{\Z}$ on Alice's and Bob's side of the
channel, respectively. In the following, we will restrict
ourselves to translational invariant channels, i.\,e., we assume
that $T$ commutes with the shift on the spin chain: $\sigma_{\A}
\circ T = T \circ \sigma_{\B}.$ In addition, we impose the
physically reasonable constraint that outputs up to some time $t$
do not depend on inputs at times $t'>t$, leading to the following

%%%%%%%%%%%%%%%%%%%%%%%%%%%%%%%%%%%%%%%%%%%%%%%%%%%%%%%%%%%%%%%%%%%%%%%%%%%%%%%%%%

\begin{define}
    \label{def:causal}
        A {\em causal channel} $T \mathpunct : \B_{\Z}
        \rightarrow \A_{\Z}$ is a completely positive and
        unital translational invariant map such that for every $z \in \Z$
        \begin{equation}
            \label{eqn:causal}
                T(b_{(-\infty,z]} \otimes \1_{[z+1,\infty)}) = T(b_{(-\infty,z]})
                \otimes \1_{[z+1,\infty)}
        \end{equation}
        for all $b_{(-\infty,z]} \in \B_{(-\infty,z]}$.
\end{define}
%%%%%%%%%%%%%%%%%%%%%%%%%%%%%%%%%%%%%%%%%%%%%%%%%%%%%%%%%%%%%%%%%%%%%%%%%%%%%%%%%%

Bearing in mind that $T$ is translational invariant, we will
henceforth set $z = 0$, and we will use the short-hands $\A_- :=
\A_{(-\infty,0]}$ and $\A_+ := \A_{[1,\infty)}$ to denote the left
and right half chain, respectively. $\B_-$ and $\B_+$ are defined
analogously.

It is obvious from the definition that a concatenated memory
channel satisfies the causality property Eq.~(\ref{eqn:causal}).
In this section we will prove the converse: every causal channel
can be represented as a concatenated memory channel. Thus, we have
the following Structure Theorem for causal channels (cf.
Fig.~\ref{fig:structure}):

%%%%%%%%%%%%%%%%%%%%%%%%%%%%%%%%%%%%%%%%%%%%%%%%%%%%%%%%%%%%%%%%%%%%%%%%%%%%%%%%%%

\begin{theo}
    \label{theo:structure}
        Let $T \mathpunct : \B_{\Z}
        \rightarrow \A_{\Z}$ be a causal channel. Ignore
        its outputs on the left half chain $\B_-$.
        Then there exists a memory observable algebra $\M$ and an
        {\em initializing channel} $R \mathpunct : \M \rightarrow
        \A_{-}$ such that $\forall \; n \in \N$
        \begin{equation}
            \label{eqn:structure}
                T(\1_{-} \otimes b_{n})
                = (R \otimes \id^{\otimes n}_{\A}) \;
                S_n (b_{n} \otimes \1_{\M})
        \end{equation}
        for all $b_n \in \B_{[1,n]} \simeq \B^{\otimes n}$, where
        $S_n$ is the $n$-fold concatenation of a memory
        channel $S \mathpunct : \B \otimes \M \rightarrow \M
        \otimes \A$, cf. Eq.~(\ref{eqn:constructive01}).
\end{theo}
%%%%%%%%%%%%%%%%%%%%%%%%%%%%%%%%%%%%%%%%%%%%%%%%%%%%%%%%%%%%%%%%%%%%%%%%%%%%%%%%%%
{\bf Proof:} In the finite-dimensional setup, a corresponding
theorem has been proved by Eggeling et al. \cite{ESW02}. Here we
generalize this result to channels on quasi-local algebras. The
Appendix contains all the background information and terminology
relevant to the proof of the theorem. As in the finite-dimensional
setting, the uniqueness of the minimal Stinespring representation
will play a crucial role.

Let $\hh$ the Hilbert space associated with the {\em universal
representation} of the left half chain $\A_-$. Note that in
general $\hh$ will not be separable. However, separability is not
required in Stinespring's Theorem. Suppose that $(\kk,\pi,V)$ is a
minimal Stinespring dilation for $T \! \mid_{\B_{-}}$, i.\,e.,
\begin{equation}
    \label{eqn:causal01}
        T(b) = V^{*} \, \pi(b) \, V^{} \quad \forall \; b \in \B_-
\end{equation}
for some Stinespring isometry $V \mathpunct : \hh \rightarrow
\kk$. In the sequel, we will make repeated use of the Hilbert
space isomorphism $\hh \simeq \hh \otimes \C_{d}^{\otimes n}$ (cf.
Ch.\:3 of Kreyszig's text \cite{Kre78}), where $\A = \B(\C_{d})$
for some $d \in \N$. From Stinespring's representation
Eq.~(\ref{eqn:causal01}) and the causality property
Eq.~(\ref{eqn:causal}), we may then conclude that
\begin{equation}
    \label{eqn:causal02}
        \begin{split}
            V^{*} \, \pi \big ( & b \otimes \1^{\otimes
            n}_{\B} \big) \, V^{}
            = T \big (b \otimes \1^{\otimes n}_{\B} \big )\\
            & = T(b) \otimes \1^{\otimes n}_{\A}\\
            & = \big ( V^{*} \otimes \1^{\otimes n}_{\A} \big ) \,
            \big ( \pi(b) \otimes \1^{\otimes n}_{\A} \big ) \, \big ( V^{}
            \otimes \1^{\otimes n}_{\A} \big)
        \end{split}
\end{equation}
for all $b \in \B_-$. Since $V$ is a minimal dilation for $T$, so
is $V \otimes \1^{\otimes n}_{\A}$ for $T \otimes \1^{\otimes
n}_{\A}$. As explained in Section~\ref{sec:stinespring} of the
Appendix, we may then conclude that there exists an isometry $W_n
\mathpunct : \kk \otimes \C_{d}^{\otimes n} \rightarrow \kk$
defined by
\begin{equation}
    \label{equation:causal02a}
        W_{n} \big ( \pi(b) \otimes \1_{\A}^{\otimes n} \big )
        \big ( V \otimes \1_{\A}^{\otimes n} \big ) \, \psi_{}
        \otimes \psi_{n} \, := \, \pi \big ( b \otimes \1_{\A}^{\otimes n} \big
        ) \, V \, \psi_{} \otimes \psi_{n}
\end{equation}
for all $b \in \B_{-}$, $ \psi_{} \in \hh$ and $\psi_{n} \in
\A^{\otimes n}$ such that
\begin{equation}
    \label{eqn:causal03}
        \pi \big ( b \otimes \1_{\B}^{\otimes n} \big ) \, W_n =
        W_n \, \big ( \pi(b) \otimes \1_{\A}^{\otimes n} \big )
\end{equation}
for all $b \in \B_-$, and
\begin{equation}
    \label{eqn:causal04}
        W_n \, \big ( V_{} \otimes \1_{\A}^{\otimes n} \big ) = V.
\end{equation}
We are now in a position to reconstruct the memory algebra: Let
$\M := \pi' (\B_-)$, the {\em commutant} of the observable algebra
$\B_-$, and let $S_n \mathpunct : \B^{\otimes n} \otimes \M
\rightarrow \bkk \otimes \B (\C_{d}^{\otimes n})$ be defined by
\begin{equation}
    \label{eqn:causal05}
        S_n (b \otimes m) := W_{n}^{*} \, \pi(b) \, m \, W_{n}^{}
\end{equation}
for all $b \in \B_{-}$ and $m \in \M$. The memory initializing
channel $R \mathpunct : \M \rightarrow \A_-$ is given by
\begin{equation}
    \label{eqn:causal06}
        R(m) := V^{*} \, m \, V^{} \quad \forall \; m \in \M.
\end{equation}
In order to justify these choices, we will first show that
\begin{equation}
    \label{eqn:causal07}
        S_n (\B^{\otimes n} \otimes \M) \subset \M \otimes
        \A^{\otimes n}.
\end{equation}
Noting that $\pi \big ( \1_{\B_{-}} \otimes \B^{\otimes n} \big )
\, \M \, \subset \pi' \big ( \B_- \otimes \1_{B}^{\otimes n} \big
)$, we see from Eq.~(\ref{eqn:causal03}) that
\begin{equation}
    \label{eqn:causal08}
        \begin{split}
            W_{n}^{*} \; & \pi  \big ( \1_{\B_-} \otimes b_n \big ) \; m
            \; W_n^{} \; \Big ( \pi \big ( \tilde{b}_{\B_-} \big ) \otimes
            \1_{A}^{\otimes n} \Big ) \\
            &= W_{n}^{*} \; \pi \big ( \1_{\B_-} \otimes b_n \big )
            \; m \; \pi \big ( \tilde{b}_{\B_-} \otimes \1_{B}^{\otimes
            n} \big ) \; W_n^{} \\
            &= W_{n}^{*} \; \pi \big ( \tilde{b}_{\B_-} \otimes \1_{B}^{\otimes
            n} \big ) \; \pi \big ( \1_{\B_-} \otimes b_n \big ) \,
            m \, W_{n}^{}\\
            &= \Big ( \pi \big ( \tilde{b}_{\B_-} \big ) \otimes
            \1_{A}^{\otimes n} \Big ) \; W_{n}^{*} \; \pi \big
            ( \1_{\B_-} \otimes b_n \big ) \; m \; W_{n}^{}
        \end{split}
\end{equation}
for all $b_n \in \B^{\otimes n}$ and $\tilde{b}_{\B_-} \in \B_-$,
implying that
\begin{equation}
    \label{eqn:causal09}
        \Big [ S_n (b_n \otimes m) \, \mid \, \pi \big (
        \tilde{b}_{\B_-} \big ) \otimes \1_{\A}^{\otimes n} \Big ]
        \, = \, 0,
\end{equation}
from which Eq.~(\ref{eqn:causal07}) directly follows. To complete
the proof, it suffices to show that $S_n$ has the right
concatenation properties, i.\,e.,
\begin{eqnarray}
    \label{eqn:causal10}
        R(m) & = & \big ( R \otimes \id_{\A}^{\otimes n} \big ) \,
        S_n \big ( \1_{\B}^{\otimes n} \otimes m \big )
        \hspace{0.5cm} {\rm and } \\
        T(b) & = & \big ( R \otimes \id_{\A}^{\otimes n} \big ) \,
        S_n \big ( b \otimes \1_{\M} \big )
\end{eqnarray}
for all $m \in \M$ and $b \in \B^{\otimes n}$. However, this is
immediate from the definitions of $S_n$ and $R$ and
Eq.~(\ref{eqn:causal04}). The result then follows by setting
$S_{} := S_{1}$. $\blacksquare$\\
%%%%%%%%%%%%%%%%%%%%%%%%%%%%%%%%%%%%%%%%%%%%%%%%%%%%%%%%%%%%%%%%%%%%%%%%%%%%%%%%%%

As can be seen from the above reasoning, the commutant algebra
$\M$ can be replaced by the von Neumann algebra generated by all
elements $( \id_{\kk} \otimes \omega_{n}) \, S_n (b_n \otimes
\1_{\M} )$. However, note that in the above construction there is
no unique way of choosing the memory algebra: given an infinite
chain of memory channels with memory algebra $\M$, considering it
as a causal channel and applying the memory reconstruction as in
the proof of Th.~\ref{theo:structure} will in general yield a
different memory algebra $\M' \neq \M$.\\

It is clear from the proof of Th.~\ref{theo:structure} that the
channel reconstruction will in general explicitly depend on the
input initializer $R$, which describes the influence of input
states in the remote past on the memory. In the following section
we will turn our attention to an important class of memory
channels for which the memory initializer becomes completely
irrelevant. These so-called {\em forgetful channels} therefore
bridge the axiomatic and the constructive approach to quantum
channels with memory. We will also show that generic memory
channels are forgetful.

%%%%%%%%%%%%%%%%%%%%%%%%%%%%%%%%%%%%%%%%%%%%%%%%%%%%%%%%%%%%%%%%%%%%%%%%%%%%%%%%%%
%%%%%%%%%%%%%%%%%%%%%%%%%%%%%%%%%%%%%%%%%%%%%%%%%%%%%%%%%%%%%%%%%%%%%%%%%%%%%%%%%%
%%%%%%%%%%%%%%%%%%%%%%%%%%%%%%%%%%%%%%%%%%%%%%%%%%%%%%%%%%%%%%%%%%%%%%%%%%%%%%%%%%
%%%%%%%%%%%%%%%%%%%%%%%%%%%%%%%%%%%%%%%%%%%%%%%%%%%%%%%%%%%%%%%%%%%%%%%%%%%%%%%%%%

\section{Forgetful Channels}
    \label{sec:forgetful}

{\em Forgetful channels} are quantum memory channels $S \mathpunct
: \B \otimes \M \rightarrow \M \otimes A$ in which the effect of
the initializing memory state dies away with time. More formally,
we have the following
%%%%%%%%%%%%%%%%%%%%%%%%%%%%%%%%%%%%%%%%%%%%%%%%%%%%%%%%%%%%%%%%%%%%%%%%%%%%%%%%%%

\begin{define}
    \label{def:forgetful}
        Let $S \mathpunct : \B \otimes \M \rightarrow \M \otimes
        \A$ be a quantum memory channel, $S_n$ its $n$-fold
        concatenation, and let $\hat{S}_n \mathpunct : \M
        \rightarrow \M \otimes \A^{\otimes n}$ be the concatenated
        channel in which Bob's outputs are ignored: $\hat{S}_n (m)
        := S_n (\1_{\B}^{\otimes n} \otimes m)$ for all $m \in
        \M$. Then $S$ is called {\em forgetful} iff there exists a
        sequence of quantum channels $\tilde{S}_n \mathpunct : \M
        \rightarrow \A^{\otimes n}$ such that
        \begin{equation}
            \label{eqn:forgetful}
                \lim_{n\to\infty} \cb{\hat{S}_n - \1_{\M} \otimes
                \tilde{S}_n} = 0.
        \end{equation}
\end{define}
%%%%%%%%%%%%%%%%%%%%%%%%%%%%%%%%%%%%%%%%%%%%%%%%%%%%%%%%%%%%%%%%%%%%%%%%%%%%%%%%%%
As an illustrative example, let's consider the classically mixed
channel $S := p \, \id \, + \, (1-p) \, S^s$, where $p \in [0,1)$,
and $S^s$ denotes the shift channel introduced in
Section~\ref{sec:examples}. When this channel is concatenated, in
every step either the ideal channel or the shift channel is chosen
with probabilities $p$ and $1-p$, respectively. The only possible
way for an $n$-fold concatenation $\hat{S}_n$ not to be forgetful
is to choose the ideal channel $\id$ in every step. However, the
probability for this event is $p^n$, and thus vanishes in the
limit $n\to\infty$, implying that Eq.\,(\ref{eqn:forgetful})
holds.
%%%%%%%%%%%%%%%%%%%%%%%%%%%%%%%%%%%%%%%%%%%%%%%%%%%%%%%%%%%%%%%%%%%%%%%%%%%%%%%%%%

\begin{remark}
    \label{remark:forgetful}
        {\em Note that Def.~\ref{def:forgetful} can be relaxed by requiring
        only that $(\tilde{S}_{n})_{n\in\N}$ is a sequence of
        linear maps, yet not necessarily channels. To see that this
        leads to an equivalent definition of forgetfulness, assume
        that $\cb{\hat{S}_n - \1_{\M} \otimes \tilde{S}_n} \leq
        \varepsilon$ for some $\varepsilon > 0$, $n\in\N$, and some
        linear operator $\tilde{S}_{n}$. Replacing
        $\1_{\M} \otimes\tilde{S}_n$ with the quantum
        channel $(P \otimes \id_{\A}^{\otimes n}) \circ
        \hat{S}_{n}$, where $P \mathpunct : \M \rightarrow \C
        \circ \1_{\M}$ is the completely depolarizing channel, we
        see that
        \begin{equation}
            \label{eqn:forgetful00}
                \begin{split}
                    \cb{\hat{S}_{n} - & (P \otimes \id_{\A}^{\otimes n})
                    \circ \hat{S}_{n}}\\
                    & \leq \cb{\hat{S}_{n} - \1_{\M} \otimes
                    \tilde{S}_n}\\
                    & \qquad +  \cb{ \big (P \otimes \id_{\A}^{\otimes
                    n} \big) \circ \big ( \1_{\M} \otimes
                    \tilde{S}_{n} - \hat{S}_{n} \big )}\\
                    & \leq \, 2 \, \cb{\hat{S}_n - \1_{\M} \otimes \tilde{S}_n}
                    \, \leq \, 2 \, \varepsilon,
                \end{split}
        \end{equation}
        and thus $\lim_{n\to\infty} \cb{\hat{S}_{n} -
        (P \otimes \id_{\A}^{\otimes n}) \circ \hat{S}_{n}} =
        0$, implying that $S$ is indeed forgetful in the sense of
        Def.~\ref{def:forgetful}.}
\end{remark}
%%%%%%%%%%%%%%%%%%%%%%%%%%%%%%%%%%%%%%%%%%%%%%%%%%%%%%%%%%%%%%%%%%%%%%%%%%%%%%%%%%
There exist several equivalent criteria for a quantum memory
channel to be forgetful. In particular, it is sufficient to show
that the norm distance $\cb{\hat{S}_n - \1_{\M} \otimes
\tilde{S}_n}$ falls below $1$ for {\em some} $n \in \N$. What is
more important, the memory effects can always be assumed to vanish
exponentially fast. In addition, if the memory algebra $\M$ has
finite dimension, the cb-norm criterion Eq.~(\ref{eqn:forgetful})
can be replaced by the usual operator norm $\norm{\cdot}$. In
fact, we have the following

%%%%%%%%%%%%%%%%%%%%%%%%%%%%%%%%%%%%%%%%%%%%%%%%%%%%%%%%%%%%%%%%%%%%%%%%%%%%%%%%%%

\begin{propo}
    \label{propo:forgetful}
        Let $S \mathpunct : \B \otimes \M \rightarrow \M \otimes
        \A$ be a quantum memory channel, and for $n\in\N$ let
        $\hat{S}_n$ be defined as in Def.~\ref{def:forgetful}.
        Then $S$ is forgetful iff there exists an integer $N \in
        \N$ and some linear operator $\tilde{S}_{N} \mathpunct :
        \M \rightarrow \A^{\otimes N}$ (not necessarily
        a channel) such that
        \begin{equation}
            \label{eqn:forgetful01}
                \cb{\hat{S}_{N} - \1_{\M} \otimes \tilde{S}_{N}} <
                1.
        \end{equation}
        Assume in addition that the memory algebra $\M$ has finite
        dimension. Then $S$ is forgetful iff for every $m\in\M$ and $\varepsilon >
        0$ we may find a positive integer $N \in\N$ and $a_{N} \in
        \A^{\otimes N}$ such that
        \begin{equation}
            \label{eqn:forgetful02}
                \norm{\hat{S}_{N} (m) - \1_{\M} \otimes a_{N}}
                \leq \varepsilon \, \norm{m}.
        \end{equation}
\end{propo}
%%%%%%%%%%%%%%%%%%%%%%%%%%%%%%%%%%%%%%%%%%%%%%%%%%%%%%%%%%%%%%%%%%%%%%%%%%%%%%%%%%
As advertised above, in the proof of Prop.~\ref{propo:forgetful}
we will also be concerned with the speed of convergence in
Eq.~(\ref{eqn:forgetful}). In this context, the following Lemma
will be helpful:

%%%%%%%%%%%%%%%%%%%%%%%%%%%%%%%%%%%%%%%%%%%%%%%%%%%%%%%%%%%%%%%%%%%%%%%%%%%%%%%%%%

\begin{lemma}
    \label{lemma:forgetful01}
        Let $(d_{n})_{n \in \N}$ be a positive and non-increasing
        sequence satisfying the subadditivity inequality
        \begin{equation}
            \label{eqn:forgetful01a}
                d_{n+m} \, \leq \, d_n \, d_m \quad \forall \; n,m
                \in \N.
        \end{equation}
        Assume further that $d_N < 1$ for some $N \in \N$. Then
        \begin{equation}
            \label{eqn:forgetful02a}
                d_{n} \, \leq \, c^n \quad \forall \; n \geq N
        \end{equation}
        for some constant $c < 1$, i.\,e., $(d_{n})_{n \in \N}$
        vanishes exponentially.
\end{lemma}
%%%%%%%%%%%%%%%%%%%%%%%%%%%%%%%%%%%%%%%%%%%%%%%%%%%%%%%%%%%%%%%%%%%%%%%%%%%%%%%%%%
{\bf Proof of Lemma~\ref{lemma:forgetful01}:} Assume that $d_N <
1$ for some $N \in \N$. From the subadditivity inequality
(\ref{eqn:forgetful01a}) we then see that $d_{N+N} \leq d_{N}^2$,
and, by induction, $d_{\nu N} \leq d_{N}^{\nu}$ for all $\nu \in
\N$. By the monotonicity of $(d_n)_{n\in\N}$ we may then conclude
that for $n \in [ \nu N, (\nu+1) N)$ we have
\begin{equation}
    \label{eqn:forgetful07}
        d_n \leq d_{\nu N} \leq d_{N}^{\nu} \leq \big (
        d_{N}^{\frac{1}{2N}} \big )^n = c^n
\end{equation}
with $c:= d_{N}^{\frac{1}{2N}} < 1$, as advertised.
$\blacksquare$\\
%%%%%%%%%%%%%%%%%%%%%%%%%%%%%%%%%%%%%%%%%%%%%%%%%%%%%%%%%%%%%%%%%%%%%%%%%%%%%%%%%%
\\
For the second part of the proof of Prop.~\ref{propo:forgetful},
we obviously need to bound the cb-norm $\cb{\cdot}$ of a linear
operator $R \mathpunct : \bhh{M} \rightarrow \A$ with $\dim
\hh_{M} < \infty$ in terms of its operator norm $\norm{\cdot}$.
This is the essence of the following
%%%%%%%%%%%%%%%%%%%%%%%%%%%%%%%%%%%%%%%%%%%%%%%%%%%%%%%%%%%%%%%%%%%%%%%%%%%%%%%%%%
\begin{lemma}
    \label{lemma:forgetful02}
        Let $R \mathpunct : \bhh{M} \rightarrow \A$ be a linear operator,
        and assume that $d_M := \dim
        \hh_{M} < \infty$. We then have
        \begin{equation}
            \label{eqn:forgetful14}
                \cb{R} \, \leq \, d_{M}^2 \, \norm{R}.
        \end{equation}
\end{lemma}
%%%%%%%%%%%%%%%%%%%%%%%%%%%%%%%%%%%%%%%%%%%%%%%%%%%%%%%%%%%%%%%%%%%%%%%%%%%%%%%%%%
{\bf Proof of Lemma~\ref{lemma:forgetful02}:} By definition of the
cb-norm, we have $\cb{R} = \sup_{k} \{ \norm{R \otimes \id_k} \}$,
where $\id_k$ is the identity operation on the $k \times k$
matrices $\B (\C_k)$. Every $x \in \bhh{M} \otimes \B (\C_k)$ can
be given the expansion
\begin{eqnarray}
    \label{eqn:forgetful15}
        x & = & \sum_{\alpha} m_{\alpha} \otimes k_{\alpha}
        \nonumber \\
        & = & \sum_{\alpha} \sum_{i,j=1}^{d_M} \mu_{\alpha,ij} \,
        \op{i}{j} \otimes k_{\alpha} \\
        & = & \sum_{i,j=1}^{d_M} \op{i}{j} \otimes x_{ij}, \nonumber
\end{eqnarray}
where we have set $x_{ij} := \sum_{\alpha} \mu_{\alpha,ij} \,
k_{\alpha}$. Note that $\norm{x_{ij}} \leq \norm{x} \; \forall \;
i,j = 1,...,d_M$, implying that
\begin{eqnarray}
    \label{eqn:forgetful16}
        \norm{ \big ( R \otimes \id_k \big ) x} & = &
        \norm{\sum_{i,j=1}^{d_M} R(\op{i}{j}) \otimes x_{ij}}
        \nonumber \\
        & \leq & \sum_{i,j=1}^{d_M} \norm{R} \, \norm{\op{i}{j}} \,
        \norm{x_{i,j}} \nonumber\\
        & \leq & d_{M}^2 \norm{R} \, \norm{x}
\end{eqnarray}
holds independently of $k$. Consequently, we have $\cb{R} =
\sup_{k} \{ \norm{R \otimes \id_k} \} \leq d_{M}^2 \norm{R}$, as
claimed. $\blacksquare$\\
%%%%%%%%%%%%%%%%%%%%%%%%%%%%%%%%%%%%%%%%%%%%%%%%%%%%%%%%%%%%%%%%%%%%%%%%%%%%%%%%%%

We now have the necessary tools at hand to tackle the\\
\\
%%%%%%%%%%%%%%%%%%%%%%%%%%%%%%%%%%%%%%%%%%%%%%%%%%%%%%%%%%%%%%%%%%%%%%%%%%%%%%%%%%
{\bf Proof of Prop.~\ref{propo:forgetful}:} We will first prove
the first part of Prop.~\ref{propo:forgetful}. Thus, at this point
we make no assumptions on the dimensionality of $\M$. If $S$ is
forgetful, Eq.~(\ref{eqn:forgetful01}) is immediate from the
definition. In order to prove the converse, let
\begin{equation}
    \label{eqn:forgetful03}
        d_n := \inf \big \{ \cb{\hat{S}_n - \1_{\M} \otimes
        \tilde{S}_{n}} \; \mid \; \tilde{S}_n \mathpunct : \M
        \rightarrow \A^{\otimes n}, {\rm linear} \big \}.
\end{equation}
for $n \in \N$. Our strategy is to show that $(d_{n})_{n\in\N}$
satisfies the conditions of Lemma~\ref{lemma:forgetful01}. From
Eq.~(\ref{eqn:forgetful01}) we can then conclude that $d_{n} \,
\leq \, c^n$ for all $n \geq N$ for some constant $c <1$, and thus
$S$ is forgetful with exponentially vanishing errors by
Remark~\ref{remark:forgetful}.

We start by showing that $(d_n)_{n\in\N}$ is non-increasing,
i.\,e., $d_{n+1} \leq d_{n} \; \forall \; n\in\N$. From the
definition of $\hat{S}_n$, we have
\begin{eqnarray}
    \label{eqn:forgetful04}
        \hat{S}_{n+1} & = & \big ( \hat{S} \otimes
        \id_{\A}^{\otimes n} \big ) \circ \hat{S}_n \nonumber \\
        & = & \big ( \hat{S} \otimes \id_{\A}^{\otimes n} \big )
        \circ \big ( \hat{S}_{n} - \1_{\M} \otimes \tilde{S}_{n}
        \big ) \nonumber \\
        & & \hspace{2cm} + \, \big ( \hat{S} \otimes \id_{\A}^{\otimes n} \big
        ) \circ \big ( \1_{\M} \otimes \tilde{S}_{n}
        \big ) \\
        & = & \big ( \hat{S} \otimes \id_{\A}^{\otimes n} \big )
        \big ( \hat{S}_{n} - \1_{\M} \otimes \tilde{S}_{n}
        \big ) + \1_{\M} \otimes \1_{\A} \otimes
        \tilde{S}_{n} \nonumber
\end{eqnarray}
where in the last step we have applied the unitality of $\hat{S}$.
From Eq.~(\ref{eqn:forgetful04}) and unitality of the cb-norm we
may conclude that
\begin{eqnarray}
    \label{eqn:forgetful05}
        d_{n+1} &\leq& \cb{\hat{S}_{n+1} - \1_{\M} \otimes \1_{\A} \otimes
        \tilde{S}_{n}} \nonumber \\
        & \leq & \cb{\hat{S} \otimes \id_{\A}^{\otimes n}} \,
        \cb{\hat{S}_{n} - \1_{\M} \otimes \tilde{S}_{n}} \leq d_n,
\end{eqnarray}
just as claimed. We will now show that $d_{n+m} \leq \, d_n \,
d_m$ for all $n,m \in \N.$ Similar to the above estimate, we have
\begin{eqnarray}
    \label{eqn:forgetful08}
        \hat{S}_{n+m} & = & \big (\hat{S}_{n} \otimes
        \id_{\A}^{\otimes m} \big ) \, \hat{S}_m \nonumber \\
        & = & \big (\hat{S}_{n} \otimes
        \id_{\A}^{\otimes m} \big ) \, \big ( \hat{S}_m - \1_{\M}
        \otimes \tilde{S}_m \big ) \nonumber \\
        & & \hspace{2cm} + \big (\hat{S}_{n} \otimes
        \id_{\A}^{\otimes m} \big ) \, \big ( \1_{\M} \otimes
        \tilde{S}_m \big ) \nonumber \\
        & = & \Big [ \big ( \hat{S}_n - \1_{\M} \otimes
        \tilde{S}_{n} \big ) \otimes \id_{\A}^{\otimes m} \Big ]
        \, \big ( \hat{S}_m - \1_{\M} \otimes \tilde{S}_m \big )
        \nonumber \\
        && \hspace{2cm} + \, \1_{\M} \otimes \tilde{S}_{n+m},
\end{eqnarray}
where we have introduced the short hand
\begin{equation}
    \label{eqn:forgetful09}
        \tilde{S}_{n+m} := \1_{\A}^{\otimes n} \otimes
        \tilde{S}_{m} + \big ( \tilde{S}_{n} \otimes
        \id_{\A}^{\otimes m} \big ) \big ( \hat{S}_m - \1_{\M} \otimes
        \tilde{S}_m \big ).
\end{equation}
Invoking again the unitality and multiplicativity of the cb-norm,
we may conclude from Eq.~(\ref{eqn:forgetful08}) that
\begin{multline}
    \label{eqn:forgetful10}
        \cb{\hat{S}_{n+m} - \1_{\M} \otimes \tilde{S}_{n+m}}\\
        \leq \cb{\hat{S}_n - \1_{\M} \otimes \tilde{S}_{n}} \,
        \cb{\hat{S}_m - \1_{\M} \otimes \tilde{S}_{m}} \leq d_{n}
        \, d_{m},
\end{multline}
which is the desired estimate. Note that $\tilde{S}_{n+m}$ is
clearly linear and unital, but not necessarily positive. This is
why we did not require the maps $\tilde{S}_{n}$ to be channels in
the definition of the sequence $(d_{n})_{n\in\N}$. This completes
the first part of the proof. $\blacktriangle$

For the second part, assume that $\M = \bhh{M}$ with $d_M := \dim
\hh_{M} < \infty$. If Eq.~(\ref{eqn:forgetful02}) holds, by the
same reasoning as in Remark~\ref{remark:forgetful} we may conclude
that $\1_{\M} \otimes a_N$ may be replaced by $(P \otimes
\id_{\A}^{\otimes N}) \circ \hat{S}_{N}(m)$, implying that for
every $m\in\M$ and $\varepsilon
> 0$ we may find a positive integer $N \in\N$ such that
\begin{equation}
    \label{eqn:forgetful12}
        \norm{\hat{S}_{N} (m) - \big (P \otimes \id_{\A}^{\otimes N}
         \big) \circ \hat{S}_{N}(m)}
         \leq 2 \, \varepsilon \, \norm{m}.
\end{equation}
In order to arrive at a uniform bound, let us introduce an
orthonormal basis $\{ \ket{i} \}_{i=1}^{d_M}$ for $\hh_{M}$. Since
$\hh_{M}$ has finite dimension, Eq.~(\ref{eqn:forgetful12}) holds
uniformally for the basis operators $\{ \op{i}{j}
\}_{i,j=1}^{d_M}$ for some possibly larger $N$. Thus, by setting
$m = \sum_{i,j = 1}^{d_M} \, m_{i,j} \, \op{i}{j}$ we see that
\begin{equation}
    \label{eqn:forgetful13}
        \begin{split}
            \norm{& \hat{S}_{N} (m)  - \big (P \otimes \id_{\A}^{\otimes N}
            \big) \circ \hat{S}_{N}(m)}\\
            & \leq \sum_{i,j=1}^{d_M} \, \abs{m_{i,j}} \,
            \norm{\hat{S}_{N} (\op{i}{j}) - \big (P \otimes \id_{\A}^{\otimes N}
            \big) \circ \hat{S}_{N}( \op{i}{j})}\\
            & \leq 2 \, \varepsilon \sum_{i,j=1}^{d_M} \, \abs{m_{i,j}}
            \, \leq \, 2 \, \varepsilon \, d_{M}^{2} \, \norm{m},
        \end{split}
\end{equation}
where in the last step we have used that $\abs{m_{i,j}} \leq
\norm{m}$ for all $i,j = 1,...,d_{M}$. Making use of
Lemma~\ref{lemma:forgetful02}, we may conclude from
Eq.~(\ref{eqn:forgetful13}) that
\begin{equation}
    \label{eqn:forgetful17}
        \cb{\hat{S}_N - \big (P \otimes \id_{\A}^{\otimes N}
         \big) \circ \hat{S}_{N}} \leq 2 \, \varepsilon \,
         d_{M}^{4}.
\end{equation}
Thus, choosing $\varepsilon < \frac{1}{2 d_{M}^4}$, we may find an
integer $N\in\N$ such that Eq.~(\ref{eqn:forgetful01}) holds.
Therefore, $S$ is forgetful by the first part of the proof. The
converse is immediate from the definition of forgetfulness.
$\blacksquare$\\
%%%%%%%%%%%%%%%%%%%%%%%%%%%%%%%%%%%%%%%%%%%%%%%%%%%%%%%%%%%%%%%%%%%%%%%%%%%%%%%%%%

From the proof of Prop.~\ref{propo:forgetful} we may immediately
deduce the following

%%%%%%%%%%%%%%%%%%%%%%%%%%%%%%%%%%%%%%%%%%%%%%%%%%%%%%%%%%%%%%%%%%%%%%%%%%%%%%%%%%

\begin{coro}
    \label{coro:speed}
        Let $S \mathpunct : \B \otimes \M \rightarrow \M \otimes
        \A$ be a forgetful quantum channel. Then the effect of the
        initial memory vanishes exponentially fast, i.\,e., we may
        find a constant $c<1$ such that
        \begin{equation}
            \label{eqn:forgetful18}
                \cb{\hat{S}_{n} - (P \otimes \id_{\A}^{\otimes n})
                \circ \hat{S}_{n}} < c^n
        \end{equation}
        for all sufficiently large $n$.
\end{coro}

%%%%%%%%%%%%%%%%%%%%%%%%%%%%%%%%%%%%%%%%%%%%%%%%%%%%%%%%%%%%%%%%%%%%%%%%%%%%%%%%%%

For convenience, and because we will use it later in
Section\,\ref{sec:codes}, in the following Proposition we show how
the definition of forgetfulness translates into the Schr\"odinger
picture language.

%%%%%%%%%%%%%%%%%%%%%%%%%%%%%%%%%%%%%%%%%%%%%%%%%%%%%%%%%%%%%%%%%%%%%%%%%%%%%%%%%%

\begin{propo}
    \label{propo:sforgetful}
        Let $S \mathpunct : \B \otimes \M \rightarrow \M \otimes
        \A$ be a quantum channel. Let $\varepsilon >0$,
        and for $n \in \N$ let $\hat{S}_n$ be defined as in
        Def.~\ref{def:forgetful}. Assume that
        \begin{equation}
            \label{eqn:forgetful19}
                \norm{\hat{S}_{n} - \big ( P \otimes \id_{\A}^{\otimes
                n} \big ) \hat{S}_{n}} \leq \varepsilon,
        \end{equation}
        where $P \mathpunct : \M \rightarrow \C \, \1_{\M}$ is a
        completely depolarizing channel. We then have
        \begin{equation}
            \label{eqn:forgetful20}
                \tracenorm{{\rm tr}_{\B^{\otimes n}} \, S_{n*} \, \big
                (\varrho_1 - \varrho_2 \big )} \leq 2 \,
                \varepsilon
        \end{equation}
        for all density operators $\varrho_{1}, \varrho_{2} \in
        \M_{*}^{} \otimes \A_{*}^{\otimes n}$ such that
        ${\rm tr}_{\M} \varrho_1 = {\rm tr}_{\M} \varrho_2$.\\ Conversely,
        suppose that Eq.\,(\ref{eqn:forgetful20}) holds. Then
        Eq.\,(\ref{eqn:forgetful19}) holds with the substitution
        $\varepsilon \mapsto 2 \, \varepsilon$.
\end{propo}

%%%%%%%%%%%%%%%%%%%%%%%%%%%%%%%%%%%%%%%%%%%%%%%%%%%%%%%%%%%%%%%%%%%%%%%%%%%%%%%%%%

In particular, if the quantum channel $S$ is forgetful, then from
Remark~\ref{remark:forgetful} we know that the condition in
Eq.\,(\ref{eqn:forgetful19}) is satisfied, and thus
Eq.\,(\ref{eqn:forgetful20}) holds. If in addition the memory
algebra $\M$ is finite-dimensional, Eq.\,(\ref{eqn:forgetful19})
is a necessary and sufficient criterion for forgetfulness by
Prop.\,\ref{propo:forgetful}. By the above Proposition,
Eq.\,(\ref{eqn:forgetful20}) then gives a necessary and sufficient
criterion for forgetfulness in the Schr\"odinger picture
language.\\
\\
%%%%%%%%%%%%%%%%%%%%%%%%%%%%%%%%%%%%%%%%%%%%%%%%%%%%%%%%%%%%%%%%%%%%%%%%%%%%%%%%%%
{\bf Proof of Prop.\,\ref{propo:sforgetful}:} Note that for any
linear operator $T \mathpunct : \B \rightarrow \A$, the operator
norm $\norm{T}$ equals the norm of the adjoint operator on the
dual space, i.\,e.,
\begin{equation}
    \label{eqn:duality}
        \norm{T} = \sup_{\tracenorm{\varrho} \leq 1}
        \tracenorm{T_{*} (\varrho)}
\end{equation}
(cf. Ch.~VI of \cite{RS80} or Section~$2.4$ of \cite{BR87} for
details). Suppose that Eq.\,(\ref{eqn:forgetful19}) holds. Since
$\id_{\A_{*}}^{\otimes n} \otimes P_{*} = {\rm tr}_{\M}$, the
partial trace on the memory algebra $\M$, we may conclude from
Eq.\,(\ref{eqn:forgetful19}) and the norm duality
Eq.\,(\ref{eqn:duality}) that
\begin{equation}
    \label{eqn:forgetful21}
        \tracenorm{\hat{S}_{n*} (\varrho) - \hat{S}_{n*} \, {\rm
        tr}_{\M} \varrho} \leq \varepsilon \; \; \forall \;
        \varrho \in \M_{*} \otimes \A_{*}^{\otimes n},
\end{equation}
which implies that for arbitrary $\varrho_{1}, \varrho_{2} \in
\M_{*} \otimes \A_{*}^{\otimes n}$ such that ${\rm tr}_{\M}
\varrho_1 = {\rm tr}_{\M} \varrho_2$ we have
\begin{equation}
    \label{eqn:forgetful22}
        \tracenorm{\hat{S}_{n*} (\varrho_1) - \hat{S}_{n*}
        (\varrho_2)} \leq 2 \varepsilon
\end{equation}
by application of the triangle inequality.
Eq.\,(\ref{eqn:forgetful20}) then follows by noting that
$\hat{S}_{n*} = {\rm tr}_{\B^{\otimes n}} \circ S_{n*}$.

Conversely, from Eq.\,(\ref{eqn:forgetful20}) we can conclude that
\begin{equation}
    \label{eqn:forgetful23}
        \tracenorm{\hat{S}_{n*} \big ( \varrho - {\rm tr}_{\M}
        \varrho \big ) } \leq 2 \, \varepsilon \; \; \forall \;
        \varrho \in \M_{*} \otimes \A_{*}^{\otimes n},
\end{equation}
which implies Eq.\,(\ref{eqn:forgetful19}) (with the substitution
$\varepsilon \mapsto 2 \, \varepsilon$) by means of the norm
duality Eq.\,(\ref{eqn:duality}). $\blacksquare$\\

%%%%%%%%%%%%%%%%%%%%%%%%%%%%%%%%%%%%%%%%%%%%%%%%%%%%%%%%%%%%%%%%%%%%%%%%%%%%%%%%%%
Prop.~\ref{propo:forgetful} (and its Schr\"odinger dual
Prop.~\ref{propo:sforgetful}) can be employed to test whether a
given quantum memory channel is forgetful. As an illustrating
example, let us consider the unitary {\em partial flip} operation
\begin{equation}
    \label{eqn:parflip}
        U_{\eta} := \cos \eta \, \F \, + \, i \sin \eta \, \1
\end{equation}
with $\eta \in [0, 2 \pi)$, where $\F := \sum_{i,j}
\op{i\,j}{j\,i}$ denotes the so-called {\em flip operator}. Since
$\F (b \otimes m) \F = m \otimes b$, for $\eta = 0$ the partial
flip is just the {\em Shift Channel} $S^{s}$ introduced in
Section~\ref{sec:examples}, which we know is forgetful. With the
help of Prop.~\ref{propo:forgetful}, we will show that the partial
flip is forgetful whenever $\cos \eta > \frac{7}{8}$. In fact, it
is sufficient to prove that
\begin{equation}
    \label{eqn:forgetful23a}
        \norm{U_{\eta} - \F} < \frac{1}{2}
\end{equation}
holds in the designated parameter range, since this will
immediately imply that
\begin{equation}
    \label{eqn:forgetful23b}
        \cb{U_{\eta}^{*} \, \1_{\B} \otimes (\cdot) \, U_{\eta}^{} -
        \F \, \1_{\B} \otimes (\cdot) \, \F} \, < \, 1,
\end{equation}
from which forgetfulness of the partial flip follows by
Prop.~\ref{propo:forgetful}. To see that
Eq.~(\ref{eqn:forgetful23a}) holds, set $\Delta_{\eta} := U_{\eta}
- \F_{}$ and observe that
\begin{equation}
    \label{eqn:forgetful23c}
        \norm{\Delta_{\eta}^{*} \, \Delta_{\eta}^{}} \, = \, 2 \, (1 -
        \cos \eta ) \,  < \, \frac{1}{4} \;
        \Longleftrightarrow \; \cos \eta > \frac{7}{8}.
\end{equation}

It seems likely that the partial flip is in fact forgetful over
the whole parameter range, apart from $\eta = \frac{1}{2} \pi$ and
$\eta = \frac{3}{2} \pi$. Evidence for this conjecture comes from
the investigation of so-called {\em collision models} by Ziman et
al. \cite{ZSB+02,SZS+02}, who could show forgetfulness of the
partial flip when the input is restricted to product states
$\varrho^{\otimes n}$.

We will prove below that forgetful quantum channels are {\em
dense} in the set of quantum memory channels: for every
non-forgetful quantum channel we may find a forgetful memory
channel which differs arbitrarily little from it. Thus, even the
partial flip at $\eta = \frac{1}{2} \pi$ and $\eta = \frac{3}{2}
\pi$ (i.\,e., the identity $\1$) can be approximated by a
forgetful quantum channel, though not necessarily a unitary one.

What is more, along the lines of the example presented above
Prop.~\ref{propo:forgetful} can be applied to show that all
quantum channels in a finite-size neighborhood of a given
forgetful quantum channel are likewise forgetful, i.\,e., the set
of forgetful quantum channels is {\em open}. Combined with the
{\em denseness} of forgetful quantum channels, this justifies the
claim made in Section~\ref{sec:outline} that generic quantum
memory channels are forgetful:
%%%%%%%%%%%%%%%%%%%%%%%%%%%%%%%%%%%%%%%%%%%%%%%%%%%%%%%%%%%%%%%%%%%%%%%%%%%%%%%%%%
\begin{theo}
    \label{theo:opendense}
        The set of forgetful quantum channels is
        open and dense in the set of quantum memory channels
        in $\cb{\cdot}$-norm topology.
\end{theo}
%%%%%%%%%%%%%%%%%%%%%%%%%%%%%%%%%%%%%%%%%%%%%%%%%%%%%%%%%%%%%%%%%%%%%%%%%%%%%%%%%%
{\bf Proof:} We will first show that the set of forgetful quantum
channels is dense in the set of quantum memory channels. From any
given (not necessarily forgetful) memory channel $S \mathpunct :
\B \otimes \M \rightarrow \M \otimes \A$ we can easily construct a
forgetful channel by mixing it with the completely depolarizing
channel
\begin{equation}
    \label{eqn:forgetful23d}
        D (b \otimes m) := \tr{(b \otimes m) \delta} \, \1_{M \otimes A},
\end{equation}
where $\delta \in \Bstar \otimes \Mstar$ is an arbitrary quantum
state. Just as in the classically mixed shift channel discussed
above, all the terms in an $n$-fold concatenation of the mixed
channel $S^{\varepsilon} := (1- \varepsilon) S^{} + \varepsilon
D^{}$ yield the identity operator $\1_{\M}$ in the memory input,
possibly apart from the $S_{n}$-contribution, which scales as $(1-
\varepsilon)^n$, and thus vanishes as $n \rightarrow \infty$.
Since this holds for all $\varepsilon > 0$, and $\cb{S^{} -
S^{\varepsilon}} \leq 2 \, \varepsilon$, we have found a forgetful
channel $S^{\varepsilon}$ arbitrarily close to $S^{}$, completing
the proof. $\blacktriangle$

We will now show that the set of forgetful quantum channels is
open. So assume that we are given a forgetful memory channel $S
\mathpunct : \B \otimes \M \rightarrow \M \otimes \A$. We will
show that $S$ has a finite-size neighborhood in which all memory
channels are forgetful. Clearly, by the definition of
forgetfulness we can find $N \in \N$ and a quantum channel
$\tilde{S}_{N} \mathpunct : \M \rightarrow \A^{\otimes N}$ such
that $\cb{\hat{S}_{N} - \1_{\M} \otimes \tilde{S}_{N}} <
\frac{1}{2}$. Thus, for all memory channels $T$ such that
$\cb{T-S} \leq \frac{1}{2 N}$ we have
\begin{equation}
    \label{eqn:forgetful24}
        \cb{\hat{T}_{N} - \1_{\M} \otimes \tilde{S}_{N}} \leq
        \cb{\hat{S}_{N} - \1_{\M} \otimes \tilde{S}_{N}} + N \,
        \cb{T - S} < 1,
\end{equation}
and the forgetfulness of $T$ immediately follows from
Prop.~\ref{propo:forgetful}. $\blacksquare$\\

%%%%%%%%%%%%%%%%%%%%%%%%%%%%%%%%%%%%%%%%%%%%%%%%%%%%%%%%%%%%%%%%%%%%%%%%%%%%%%%%%%
It is instructive to observe that a forgetful channel is obtained
from a possibly non-forgetful one in the denseness proof of
Th.~\ref{theo:opendense} by adding a tiny amount of white noise.
In real-world experiments, such noise will always be present at
some level. Therefore, quantum channels encountered in the
laboratory will generally be forgetful.

However, while every non-forgetful quantum channel can be
approximated by a forgetful memory channel to arbitrary degree of
accuracy, their capacities may be different. As an example for
such a discontinuity effect, consider the channel with a global
classical switch introduced in Section~\ref{sec:examples}. Let us
assume that Alice and Bob face a situation in which Eve controls
the initial memory state and completely jams the communication.
Then adding a little bit of noise, as in the proof of
Th.~\ref{theo:opendense}, will deprive Eve of her control of the
initial memory, and may lead to a channel with positive
transmission rate. Thus, adding noise may actually be beneficial
sometimes. Of course, it is just as easy to construct examples of
memory channels which are rendered useless by adding a tiny amount
of noise.\\
\\
In the special case of unitary quantum channels asymptotically
vanishing memory effects have been investigated by Wellens et al.
\cite{WBK+00} under the name {\em asymptotic completeness}, with a
special focus on the preparation of arbitrary memory output
states. While asymptotic completeness and forgetfulness are
certainly related concepts, they seem to differ in fine points,
for instance in the choice of the operator topology. Asymptotic
completeness of the Jaynes-Cummings interaction, which governs the
physics of the micromaser experiment described in
Section~\ref{sec:related}, is claimed as a main mathematical
result in \cite{WBK+00}. However, a proof is neither available in
the cited literature \cite{KM00}, nor upon request \cite{Kue05}.

%%%%%%%%%%%%%%%%%%%%%%%%%%%%%%%%%%%%%%%%%%%%%%%%%%%%%%%%%%%%%%%%%%%%%%%%%%%%%%%%%%
%%%%%%%%%%%%%%%%%%%%%%%%%%%%%%%%%%%%%%%%%%%%%%%%%%%%%%%%%%%%%%%%%%%%%%%%%%%%%%%%%%
%%%%%%%%%%%%%%%%%%%%%%%%%%%%%%%%%%%%%%%%%%%%%%%%%%%%%%%%%%%%%%%%%%%%%%%%%%%%%%%%%%
%%%%%%%%%%%%%%%%%%%%%%%%%%%%%%%%%%%%%%%%%%%%%%%%%%%%%%%%%%%%%%%%%%%%%%%%%%%%%%%%%%

\section{Entropic Bounds and Channel Coding}
    \label{sec:codes}

While in Section~\ref{sec:examples} and Section~\ref{sec:pure} we
have computed the channel capacity of some interesting model
channels, in this section we will be concerned with statements
that apply more generally. In Section~\ref{sec:bounds} we will
give entropic upper bounds on the capacity for classical and
quantum information transfer. In Section~\ref{sec:coding}
achievability of these bounds will be demonstrated for forgetful
quantum channels.

%%%%%%%%%%%%%%%%%%%%%%%%%%%%%%%%%%%%%%%%%%%%%%%%%%%%%%%%%%%%%%%%%%%%%%%%%%%%%%%%%%
%%%%%%%%%%%%%%%%%%%%%%%%%%%%%%%%%%%%%%%%%%%%%%%%%%%%%%%%%%%%%%%%%%%%%%%%%%%%%%%%%%

\subsection{Entropic Bounds}
    \label{sec:bounds}

It has already been pointed out by Bowen and Mancini \cite{BM03}
that the standard mutual information bound (or {\em Holevo bound})
\cite{Hol73} on the classical channel capacity as well as the
coherent information bound \cite{BNS98,BST98,BKN00,Dev03} on the
quantum capacity can be extended to quantum channels with memory.
In fact, these bounds ultimately depend only on the mutual
information between Alice's input register and Bob's output
register, and are independent of the internal structure of the
quantum channel that links both parties. The proofs familiar from
the memoryless setting can therefore be directly applied to memory
channels, and yield entropic upper bounds on the classical and
quantum capacity of a quantum memory channel in all the four
different settings discussed in Def.~\ref{def:cap}.

Before we state these bounds in Props.~\ref{propo:cbounds} and
\ref{propo:qbounds} below, we will need to introduce some notation
and terminology. In the following, the {\em von Neumann entropy}
of a quantum state $\varrho \in \bhstar$ will be denoted by
$H(\varrho) := - \tr{\varrho \, \ld \varrho}$. Given a quantum
channel (in Schr\"odinger picture) $S_{*} \mathpunct : \bhhstar{1}
\rightarrow \bhhstar{2}$ and an ensemble $\{p_i, \varrho_i
\}_{i=1}^{I}$ of quantum states $\varrho_i \in \bhhstar{1}$, where
$\{ p_i \}_{i=1}^{I}$ is a classical probability distribution,
Holevo's {\em $\chi$-quantity} is given by
\begin{equation}
    \label{eqn:bounds01}
        \chi(S_{*}, \{p_i, \varrho_i \}) := H \big (
        \sum_{i=1}^{I} p_i \, S_{*}(\varrho_i) \big ) -
        \sum_{i=1}^{I} p_i \, H \big ( S_{*}(\varrho_i) \big ).
\end{equation}
The {\em coherent information} $I_c (S_{*}, \varrho)$ of the
quantum channel $S_*$ with respect to a state $\varrho \in
\bhhstar{1}$ is likewise given in terms of the von Neumann
entropy,
\begin{equation}
    \label{eqn:bounds02}
        I_c (S_{*}, \varrho) := H \big ( S_{*}(\varrho) \big ) - H
        \big ( S_* \otimes \id (\kb{\psi}) \big ),
\end{equation}
where $\psi \in \hh_1 \otimes \hh_1$ is a {\em purification} of
the quantum state $\varrho \in \bhhstar{1}$ \cite{NC00}. With
these notations, we have the following
%%%%%%%%%%%%%%%%%%%%%%%%%%%%%%%%%%%%%%%%%%%%%%%%%%%%%%%%%%%%%%%%%%%%%%%%%%%%%%%%%%
\begin{propo}
    \label{propo:cbounds}
    Let $S_{n*}$ be the $n$-fold concatenation of a quantum memory
    channel $S_{*} \mathpunct : \bhhstar{M} \otimes \bhhstar{A} \rightarrow
    \bhhstar{B} \otimes \bhhstar{M}$. The classical information capacities of $S$
    are bounded from above as follows:
    \begin{align}
        \label{eqn:bounds03}
            C_{AB}(S) & \leq \varlimsup_{n\to\infty}
            \frac{1}{n} \max_{\{ p_i,\varrho_i \}} \chi(S_{n*}, \{p_i, \varrho_i
            \}),\\
        \label{eqn:bounds04}
            C_{AE}(S) & \leq \varlimsup_{n\to\infty}
            \frac{1}{n} \max_{\{ p_i,\varrho_i \}} \chi({\rm tr}_{\M} \circ
            S_{n*}, \{p_i, \varrho_i\}),\\
        \label{eqn:bounds05}
            C_{EB,\mu}(S) & \leq \varlimsup_{n\to\infty}
            \frac{1}{n} \max_{\{ p_i,\varrho_i \}} \chi(
            S_{n*}, \{p_i, \mu \otimes \varrho_i \}),\\
        \label{eqn:bounds06}
            C_{EE,\mu}(S) & \leq \varlimsup_{n\to\infty}
            \frac{1}{n} \max_{\{ p_i,\varrho_i \}} \chi({\rm tr}_{\M} \circ
            S_{n*}, \{p_i, \mu \otimes \varrho_i \}),
    \end{align}
    where $\mu \in \bhhstar{M}$ is Eve's initial memory state. If
    $d_{M} := \dim \hh_{M} < \infty$, the bounds in
    Eq.~(\ref{eqn:bounds03}), Eq.~(\ref{eqn:bounds04}) and in
    Eq.~(\ref{eqn:bounds05}), Eq.~(\ref{eqn:bounds06}) coincide
    pairwise. If the channel $S$ is forgetful, the bounds in
    Eq.~(\ref{eqn:bounds03}), Eq.~(\ref{eqn:bounds05}) and in
    Eq.~(\ref{eqn:bounds04}), Eq.~(\ref{eqn:bounds06}) coincide
    pairwise.
\end{propo}
%%%%%%%%%%%%%%%%%%%%%%%%%%%%%%%%%%%%%%%%%%%%%%%%%%%%%%%%%%%%%%%%%%%%%%%%%%%%%%%%%%

\begin{propo}
    \label{propo:qbounds}
    The quantum information capacities of the memory channel $S$
    are bounded from above as follows:
    \begin{align}
        \label{eqn:bounds07}
            Q_{AB}(S) & \leq \varlimsup_{n\to\infty}
            \frac{1}{n} \max_{\varrho} I_{c} (S_{n*}, \varrho), \\
        \label{eqn:bounds08}
            Q_{AE}(S) & \leq \varlimsup_{n\to\infty}
            \frac{1}{n} \max_{\varrho} I_{c} ({\rm tr}_{\M} \circ S_{n*}, \varrho), \\
        \label{eqn:bounds09}
            Q_{EB,\mu}(S) & \leq \varlimsup_{n\to\infty}
            \frac{1}{n} \max_{\varrho} I_{c} (S_{n*}, \mu \otimes \varrho), \\
        \label{eqn:bounds10}
            Q_{EE,\mu}(S) & \leq \varlimsup_{n\to\infty}
            \frac{1}{n} \max_{\varrho} I_{c} ({\rm tr}_{\M} \circ S_{n*},
            \mu \otimes \varrho),
    \end{align}
    where $\mu \in \bhhstar{M}$ is Eve's initial memory state.
    If $d_{M} < \infty$, the bounds in
    Eq.~(\ref{eqn:bounds07}), Eq.~(\ref{eqn:bounds08}) and in
    Eq.~(\ref{eqn:bounds09}), Eq.~(\ref{eqn:bounds10}) coincide
    pairwise. If the channel $S$ is forgetful, the bounds in
    Eq.~(\ref{eqn:bounds07}), Eq.~(\ref{eqn:bounds09}) and in
    Eq.~(\ref{eqn:bounds08}), Eq.~(\ref{eqn:bounds10}) coincide
    pairwise.
\end{propo}
%%%%%%%%%%%%%%%%%%%%%%%%%%%%%%%%%%%%%%%%%%%%%%%%%%%%%%%%%%%%%%%%%%%%%%%%%%%%%%%%%%

\begin{remark}
    \label{remark:optimistic}
    {\em Note that the bounds in Props.~\ref{propo:cbounds} and
    \ref{propo:qbounds} still hold when we only require that
    coding is possible along some (possibly very
    sparse) block sequence $(n_\nu)_{\nu\in\N}$. In Def.~\ref{def:cap} we have been more
    ambitious, since we have required that coding works for arbitrary block
    size. When this stronger version of capacity is chosen, the $\varlimsup$ can be
    replaced by $\varliminf$ in Eqs.~(\ref{eqn:bounds03}) through
    (\ref{eqn:bounds10}). While the ``optimistic" and the
    ``pessimistic" channel capacity coincide for memoryless channels
    \cite{KW04}, this is not clear for channels with memory (cf.
    Remark~\ref{remark:equivalent}). For forgetful channels,
    equivalence does hold, as will be seen in
    Section~\ref{sec:coding}.}
\end{remark}
%%%%%%%%%%%%%%%%%%%%%%%%%%%%%%%%%%%%%%%%%%%%%%%%%%%%%%%%%%%%%%%%%%%%%%%%%%%%%%%%%%
{\bf Proof of Props.~\ref{propo:cbounds} and \ref{propo:qbounds}:}
As indicated above, the proof transfers directly from the
memoryless setting. We thus refer to Holevo's original work
\cite{Hol73} for the classical bound, and to the works of Barnum
et al. \cite{BNS98,BST98,BKN00} and Devetak \cite{Dev03} for the
quantum case.

Here we only show that the bounds coincide pairwise under the
additional assumption of having a memory of finite size or a
forgetful channel. We will begin with the finite memory case: Note
that the Holevo quantity $\chi$ decreases under quantum
operations, i.\,e.,
\begin{equation}
    \label{eqn:bounds11}
        \chi(R_{*} \, S_{*}, \{p_i, \varrho_i \})
        \leq \chi(S_{*}, \{p_i, \varrho_i \})
\end{equation}
for any pair of quantum channels $R_{*}, S_{*}$ and any ensemble
of quantum states $\{ p_i, \varrho_i \}_i$ \cite{NC00}. We see
from Eq.~(\ref{eqn:bounds11}) that
\begin{multline}
    \label{eqn:bounds12}
        \chi({\rm tr}_{\M} \circ S_{n*}, \{p_i, \varrho_i \}) \leq
        \chi(S_{n*}, \{p_i, \varrho_i \}) \\
        \leq \chi({\rm tr}_{\M} \circ S_{n*}, \{p_i, \varrho_i \})
        + 2 \, \ld d_{M},
\end{multline}
where in the last step the subadditivity of von Neumann entropy
has been applied \cite{NC00}. From Eq.\,(\ref{eqn:bounds12}) it
immediately follows that the bounds on $C_{AB}$ and $C_{AE}$
coincide whenever $d_{M} < \infty$. The proof for the bounds on
$C_{EB,\mu}$ and $C_{EE,\mu}$ is completely analogous.

For the bounds on the quantum capacities, replace
Eq.\,(\ref{eqn:bounds11}) by the {\em Data Processing Inequality},
i.\,e.,
\begin{equation}
    \label{eqn:DPI}
        I_c ( R_* \circ S_*, \varrho) \leq I_c (S_*, \varrho)
\end{equation}
for any two quantum channels $R_*$ and $S_*$ \cite{NC00}, and
again apply subadditivity of von Neumann entropy. $\blacktriangle$

In the forgetful setting, in addition to subadditivity of von
Neumann entropy we will also need to make use of its continuity
properties. In fact, by {\em Fannes' Inequality} \cite{Fan73,NC00}
we have
\begin{equation}
    \label{eqn:fannes}
        \abs{H(\varrho) - H(\sigma)} \leq \tracenorm{\varrho -
        \sigma} \, \ld d \, + \, \frac{\ld e}{e},
\end{equation}
where $\varrho, \sigma \in \bhstar$ are quantum states, and $d :=
\dim \hh$.

By the results of Prop.\,\ref{propo:sforgetful}, forgetfulness of
the channel $S$ implies that for any $\varepsilon
> 0$ we may find a positive integer $m\in\N$ such that
\begin{equation}
    \label{eqn:bounds12a}
        \tracenorm{{\rm tr}_{\B^{\otimes m}} \, S_{m*} \, \big (\varrho_1
        - \varrho_2 \big )} \leq \, \varepsilon
\end{equation}
for all density operators $\varrho_{1}, \varrho_{2} \in
\bhhstar{M} \otimes \bhhstar{A}^{\otimes n}$ satisfying ${\rm
tr}_{\M} \varrho_1 = {\rm tr}_{\M} \varrho_2$. Applying Fannes'
Inequality Eq.\,(\ref{eqn:fannes}) and subadditivity of von
Neumann entropy, we can thus conclude that for arbitrary $\mu \in
\bhhstar{M}$ and $n\in\N$ we have
\begin{equation}
    \label{eqn:bounds13}
        \begin{split}
            \chi & (S_{n*}, \{p_i, \varrho_i \})  \leq
            \chi({\rm tr}_{\B^{\otimes m}} \, S_{n*},
            \{p_i, \varrho_i \}) + 2 \, m \, \ld d_B\\
            & \leq \chi({\rm tr}_{\B^{\otimes m}} \, S_{n*},
            \{p_i, \mu \otimes{\rm tr}_{\M} (\varrho_i) \})
            + 2 \, m \, \ld d_B\\
            & \quad \quad + \frac{2 \, \ld e}{e} +
            2 \, \tracenorm{{\rm tr}_{\B^{\otimes m}}
            S_{n*} \big ( \varrho_i - \mu \otimes {\rm tr}_{\M} ( \varrho_i )
            \big ) } \, \ld d_{B}^{n}\\
            & \leq \max_{\{q_j, \sigma_j\}}
            \chi(S_{n*}, \{q_j, \mu \otimes \sigma_j \})
            + 2 \, m \, \ld d_B \\
            & \quad \quad  + \frac{2 \, \ld e}{ e} +
            2 \, n \, \varepsilon \, \ld d_{B}.
        \end{split}
\end{equation}
Maximizing over the ensemble $\{p_i, \varrho_i \}$, dividing by
$n$ and letting $n\to\infty$, we may conclude from
Eq.\,(\ref{eqn:bounds13}) that
\begin{multline}
    \label{eqn:bounds14}
        \varlimsup_{n\to\infty} \frac{1}{n} \max_{\{p_i, \varrho_i \}}
         \chi(S_{n*}, \{p_i, \varrho_i \}) \\ \leq
         \varlimsup_{n\to\infty} \frac{1}{n} \max_{\{p_i, \varrho_i \}}
         \chi(S_{n*}, \{p_i, \mu \otimes \varrho_i \}) + 2 \,
         \varepsilon \, \ld d_{B},
\end{multline}
implying that for every $\mu \in \bhhstar{M}$ the bound on the
classical capacity $C_{EB,\mu}$ is no smaller than the bound on
the capacity $C_{AB}$. The converse estimate is immediate, since
Alice can obviously choose quantum ensembles of the form $\{p_i,
\mu \otimes \varrho_i \}$ if she has access to the input memory.
The proof for the bounds on $C_{EE,\mu}$ and $C_{AE}$ is
completely analogous, as is the proof for the quantum case.
$\blacksquare$

%%%%%%%%%%%%%%%%%%%%%%%%%%%%%%%%%%%%%%%%%%%%%%%%%%%%%%%%%%%%%%%%%%%%%%%%%%%%%%%%%%
%%%%%%%%%%%%%%%%%%%%%%%%%%%%%%%%%%%%%%%%%%%%%%%%%%%%%%%%%%%%%%%%%%%%%%%%%%%%%%%%%%

\subsection{Coding Theorems for Forgetful Channels}
    \label{sec:coding}

In this section we will demonstrate that for forgetful channels
the entropic bounds on the classical and quantum channel
capacities presented in Prop.\,\ref{propo:cbounds} and
Prop.\,\ref{propo:qbounds} are in fact achievable rates, and the
limits exist.

The idea of the proof is a reduction of the problem to the
memoryless setting via a relatively simple double-blocking
procedure. To illustrate the strategy, let's start with the easy
case in which there is a finite integer $m\in\N$ such that
\begin{equation}
    \label{eqn:coding03}
        \hat{S}_{m} = (P \otimes \id_{\A}^{\otimes m})
        \circ \hat{S}_{m},
\end{equation}
where $P \mathpunct : \M \rightarrow \C \, \1_{\M}$ is again the
completely depolarizing channel. We call channels with this
property {\em strictly forgetful}, and the smallest integer $m$
such that Eq.~(\ref{eqn:coding03}) is satisfied will be called the
{\em memory depth} of the channel $S$. For the processing of long
messages, we group the channels into blocks of length $m+l$ and
ignore the outputs of the first $m$ channels of each block, while
the actual coding is done for the remaining $l$ channels.
Eventually we will let $l\to\infty$. When we restrict the inputs
to product states of block length $m+l$, due to strict
forgetfulness the output state factorizes, and the whole setup
corresponds to a memoryless channel on the larger input space
$\hh^{\otimes l+m}_{A}$. For the transmission of classical
information, we can then apply the standard random coding
techniques of Holevo \cite{Hol98} and Schumacher and Westmoreland
\cite{SW97}. Invoking subadditivity of von Neumann entropy as in
Section~\ref{sec:bounds}, the rates $R$ which can be achieved with
this coding scheme are seen to be bounded as follows:
\begin{multline}
    \label{eqn:coding04}
        \frac{1}{l+m} \max_{\{ p_i,\varrho_i \}} \chi(S_{l*}, \{p_i,
        \varrho_i\}) - \frac{2\, m}{m+l} \, \ld d_B \\\leq R \leq
        \frac{1}{l} \max_{\{ p_i,\varrho_i \}} \chi(S_{l*}, \{p_i,
        \varrho_i\}).
\end{multline}
The claim then follows by letting $l\to\infty$. For quantum
channel capacities, Devetak's coding theorem \cite{Dev03} can be
shown to yield an analogous bound, in which the Holevo quantity is
replaced by coherent information.

It turns out that we can apply the same double-blocking strategy
even if the memory channel $S$ is merely assumed to be forgetful
(and no longer strictly forgetful). However, in this case the
output does not completely factorize, and the error we pick up by
replacing the memory channel with a memoryless channel on larger
blocks grows with the number of blocks. Luckily, all memory
effects can be assumed to vanish exponentially fast by
Corollary~\ref{coro:speed}.\\

While in this paper we have focused on the classical and quantum
channel capacities proper, Devetak's proof of the quantum channel
coding theorem \cite{Dev03} is based on a coherentification scheme
for the {\em private} classical channel capacity. The setup for
private information transfer (including the definition of rates
and capacity) is almost the same as for classical channel
capacity, but the protocols have to satisfy the additional
requirement that (almost) no information is released to the
environment.

More formally, assume that a quantum channel $T_{*} \mathpunct :
\bhhstar{A} \rightarrow \bhhstar{B}$ is implemented by the
Stinespring isometry $V \mathpunct : \hh_{A} \rightarrow \hh_{B}
\otimes \hh_{E}$, i.\,e.,
\begin{equation}
    \label{eqn:coding14}
        T_{*} (\varrho) = {\rm tr_{E}} \; V^{} \varrho V^{*} \; \;
        \forall \; \varrho \in \bhhstar{A}
\end{equation}
(cf. Section~\ref{sec:stinespring} of the Appendix for details).
By $T_{*}^{E}$ we then denote the channel that arises from $T_{*}$
by interchanging the roles of $\hh_B$ and $\hh_E$, i.\,e.,
\begin{equation}
    \label{eqn:coding15}
        T_{*}^{E} (\varrho) := {\rm tr_{B}} \; V^{} \varrho V^{*} \; \;
        \forall \; \varrho \in \bhhstar{A}.
\end{equation}
This channel describes the information flow into the environment.
Privacy in Devetak's coding scheme for memoryless channels then
means that for sufficiently large $n \in \N$ we may find an
operator $\Theta \in \bhh{E}^{\otimes n}$  such that
\begin{equation}
    \label{eqn:coding15a}
        \Big \| \frac{1}{\nu_{E}} \sum_{k=1}^{\nu_{E}}
        \, T^{E \; \otimes n}_{*}
        \left (\varrho_{jk} \right) \, - \, \Theta   \Big
        \|_{1}
        \leq \varepsilon \; \; \forall \; j=1,...,\nu_{B},
\end{equation}
where $\{ \varrho_{jk} \}_{j=1,k=1}^{\nu_{B}, \nu_{E}}$ is a set
of codewords, and $\nu_{B} = 2^{nR}$ describes the size of the
code space necessary to attain the rate $R>0$. We see from
Eq.~(\ref{eqn:coding15a}) that privacy is achieved by randomizing
over part of the codewords, leading to smaller code spaces.
Devetak could show \cite{Dev03} that the capacity $C^{p}(T)$ of a
memoryless quantum channel $T$ for private classical information
transfer is given by
\begin{equation}
    \label{eqn:coding16a}
        \begin{split}
            C^{p}(T) = \lim_{n\to\infty} \frac{1}{n} \max_{\{ p_{i}, \varrho_{i}
            \}} & \big \{ \chi(T^{\otimes n}, \{ p_{i}, \varrho_{i}
            \} )\\
            & \quad -  \chi(T^{E \; \otimes n}, \{ p_{i}, \varrho_{i}
            \} ) \big \},
        \end{split}
\end{equation}
where $\chi$ is the Holevo quantity introduced in
Eq.~(\ref{eqn:bounds01}).

It is a coherent version of this private classical information
protocol which yields the quantum channel coding theorem. Note in
particular that if $\varrho = \sum_i \, p_i \kb{\psi_i}$ is a
decomposition of $\varrho \in \bhhstar{A}$ into pure states, we
have
\begin{equation}
    \label{eqn:coding16b}
        I_c (T^{}, \varrho) = \chi(T^{}, \{ p_{i}, \kb{\psi_i}
        \} ) \, - \, \chi(T^{E}, \{ p_{i}, \kb{\psi_i}
        \} )
\end{equation}
by the Joint Entropy Theorem (cf. Th. 11.8 of \cite{NC00}).\\

As described above, part of our strategy in this Section will be
an extension of Devetak's coherentification protocol to forgetful
quantum channels. In fact, the coherentification protocol itself
applies generally and does not depend on the internal structure of
the quantum channel that links the sender to the receiver and the
environment. Thus, our proof of the quantum coding theorem amounts
to showing that the privacy condition Eq.~(\ref{eqn:coding15a})
can be satisfied for forgetful quantum channels. Consequently, in
the course of the proof we will also obtain a coding theorem for
the private classical information of forgetful quantum channels.
We thus have the following

%%%%%%%%%%%%%%%%%%%%%%%%%%%%%%%%%%%%%%%%%%%%%%%%%%%%%%%%%%%%%%%%%%%%%%%%%%%%%%%%%%
\begin{theo}
    \label{theo:coding}
    Let $\hh_{A}$, $\hh_{B}$, and $\hh_{M}$ be finite-dimensional
    Hilbert spaces, and let us assume that $S_{*} \mathpunct : \bhhstar{M}
    \otimes \bhhstar{A} \rightarrow \bhhstar{B} \otimes
    \bhhstar{M}$ is a forgetful quantum channel. By $S_{n*}$ we denote its
    $n$-fold concatenation. With the convention introduced in
    Remark~\ref{remark:star}, we then have
    \begin{align}
        \label{eqn:coding01}
            C_{*}(S) & =  \lim_{n\to\infty}
            \frac{1}{n} \max_{\{ p_i,\varrho_i \}} \chi(S_{n*}, \{p_i, \varrho_i
            \}),\\
        \label{eqn:coding01a}
            C_{*}^{p}(S) & =  \lim_{n\to\infty}
            \frac{1}{n} \max_{\{ p_i,\varrho_i \}}  \chi(S_{n*}^{}, \{p_i, \varrho_i
            \}) - \chi(S_{n*}^{E}, \{p_i, \varrho_i \}),\\
        \label{eqn:coding02}
             Q_{*}(S) & =  \lim_{n\to\infty}
            \frac{1}{n} \max_{\varrho} I_{c} (S_{n*}, \varrho).
    \end{align}
\end{theo}
%%%%%%%%%%%%%%%%%%%%%%%%%%%%%%%%%%%%%%%%%%%%%%%%%%%%%%%%%%%%%%%%%%%%%%%%%%%%%%%%%%
{\bf Proof:} The proof of the upper bound on the private classical
capacity $C_{*}^{p}(S)$, i.\,e.,
\begin{equation}
    \label{eqn:coding02b}
        C_{AB}^{p}(S) \leq  \varlimsup_{n\to\infty}
        \frac{1}{n} \max_{\{ p_i,\varrho_i \}}  \chi(S_{n*}^{}, \{p_i, \varrho_i
        \}) - \chi(S_{n*}^{E}, \{p_i, \varrho_i \}),
\end{equation}
is completely analogous to the one for the memoryless case
\cite{Dev03}. For $C_{AB}(S)$ and $Q_{AB}(S)$, corresponding
results have been presented in Props.~\ref{propo:cbounds} and
\ref{propo:qbounds}. To complete the proof it thus remains to show
that
\begin{equation}
    \label{eqn:coding05}
        C_{EE,\mu}(S)  \geq  \lim_{n\to\infty}
            \frac{1}{n} \max_{\{ p_i,\varrho_i \}} \chi(S_{n*}, \{p_i,
            \varrho_i \})
\end{equation}
for all $\mu \in \bhhstar{M}$, and that the limit on the right
hand side of Eq.~(\ref{eqn:coding05}) exists, and correspondingly
for $C^{p}_{EE,\mu}(S)$ and $Q_{EE,\mu}(S)$.

The definition of forgetfulness combined with
Corollary\,\ref{coro:speed} implies that we may find a sequence
$(\tilde{S}_{m})_{m\in\N}$ of quantum channels such that
\begin{equation}
    \label{eqn:coding06}
        \cb{\hat{S}_{m} - \1_{\M} \otimes \tilde{S}_{m}} \leq
        c^{-m}
\end{equation}
for some constant $c>1$.

As described above for the case of strictly forgetful channels,
our strategy is then to group the memory channels into blocks of
length $m+l$, to ignore the outputs on the first $m$ channels of
each block, and to replace the resulting channel $T_{m+l} :=
(\hat{S}_m \otimes \id_{\A^{\otimes l}}) \circ S_{l}$ by the
memoryless channel
\begin{equation}
    \label{eqn:coding06a}
        \tilde{T}_{m+l} := ( \1_{\M} \otimes \tilde{S}_m \otimes
        \id_{\A^{\otimes l}}) \circ S_{l}.
\end{equation}
For Alice, this coding procedure means that she will have to feed
the first $m$ inputs of each block of length $m+l$ with some
standard state $\omega \in \bhhstar{A}^{\otimes m}$, while she
will use the remaining $l$ inputs of each block for the actual
coding. Bob will ignore the first $m$ output signals of each
block, and will run his decoding algorithm on the remaining $l$
signals.

Let us focus on the classical information capacity first, and
assume that we have a coding scheme for the memoryless channel
$\tilde{T}_{m+l}$ that achieves the rate $R \in \R$. By definition
of capacity, this means that for every $\varepsilon > 0$ there is
an integer $N_{\varepsilon} \in \N$ such that for every $n\geq
N_{\varepsilon}$ we may find a code book with $\nu := \lfloor 2^{n
l R} \rfloor$ codewords $\{ \varrho_j \}_{j=1}^{\nu} \subset
\bhhstar{A}^{\otimes l n}$ and a corresponding observable $\{ M_j
\}_{j=1}^{\nu} \subset \bhh{B}^{\otimes l n}$ such that
\begin{equation}
    \label{eqn:coding07}
        {\rm tr} \, \tilde{T}_{m+l*}^{\otimes n}
        (\varrho_j) M_j \; \geq \; 1 - \varepsilon \; \; \forall \; n \geq
        N_{\varepsilon},
\end{equation}
uniformly in $\{ \varrho_j \}_{j=1}^{\nu}$. By the results of
Holevo \cite{Hol98} and Schumacher and Westmoreland \cite{SW97},
such coding schemes exist for all rates $R < \frac{l}{m+l} C_1
(\tilde{T}_{l})$, where $C_1 (\tilde{T}_{l})$ denotes the product
state capacity of the memoryless channel $\tilde{T}_{l}$.

For the private classical information capacity, the setting is
basically the same, but the codewords $\{ \varrho_{jk}
\}_{j=1,k=1}^{\nu_{B}, \nu_{E}}$ carry a second index to allow for
randomization, and there exists an operator $\Theta \in
\bhh{E}^{\otimes nl}$ such that
\begin{equation}
    \label{eqn:coding07a}
        \Big \| \frac{1}{\nu_{E}} \sum_{k=1}^{\nu_{E}}
        \tilde{T}^{E \; \otimes n}_{l*}
        \left ( \varrho_{jk} \right) \, - \, \Theta   \Big
        \|_{1}
        \leq \varepsilon \quad \forall \; j=1,...,\nu_{B}
\end{equation}
(cf. Eq.~(\ref{eqn:coding15a}) above). Here the size of the code
is given by $\nu_{B} = \lfloor 2^{nlR} \rfloor$, and all rates $R
< \frac{l}{l+m} C_{1}^{p}(\tilde{T}_{l})$ may be achieved.

The same product coding scheme will now be applied to the
concatenated memory channel $T_{m+l}$. Our objectives are to show
that
\begin{enumerate}
    \renewcommand{\labelenumi}{$(\alph{enumi})$}
    \item this coding scheme satisfies the decoding condition
        Eq.~(\ref{eqn:coding07}),
    \item in the case of private information transfer,
        the privacy condition Eq.~(\ref{eqn:coding07a}) holds, and
    \item the attainable rates can be made arbitrarily close to the
        entropic upper bounds.
\end{enumerate}

This will immediately imply the coding theorem for classical and
private classical information transfer. The quantum channel coding
theorem will then follow from the coherentification of the private
classical protocol, as explained in detail in Devetak's original
work \cite{Dev03}.

Let us start with the decoding condition $(a)$. Assume that in $n$
blocks of length $m+l$ each, the replacement $T_{m+l} \mapsto
\tilde{T}_{m+l}$ is made. Since $\cb{T_{m+l} - \tilde{T}_{m+l}}
\leq c^{-m}$ for each of these blocks by Eq.~(\ref{eqn:coding06}),
the concatenated channels satisfy
\begin{equation}
    \label{eqn:coding07b}
        \cb{T_{n(m+l)}^{} - \tilde{T}_{m+l}^{\otimes n}} \, \leq \, n \,
        c^{-m}.
\end{equation}
Making use of the norm duality Eq.\,(\ref{eqn:duality}), we can
conclude from Eq.\,(\ref{eqn:coding07b}) that
\begin{equation}
    \label{eqn:coding08}
            \tracenorm{T_{n(m+l)*}  (\varrho) -
            \tilde{T}_{m+l*}^{\otimes n} (\varrho)} \leq n \,
            c^{-m}.
\end{equation}
Noting that for any two quantum states $\varrho, \sigma \in
\bhstar$ and any observable $\{ M_j \}_{j=1}^{\nu} \subset \bh$
the inequality
\begin{equation}
    \label{eqn:coding09}
        \tracenorm{\varrho - \sigma} \geq \sum_{j=1}^{\nu}
        \abs{{\rm tr} M_j (\varrho - \sigma)}
\end{equation}
holds (cf. Th. 9.1 of \cite{NC00}), we may infer from
Eq.\,(\ref{eqn:coding08}) that for all codewords $\{ \varrho_j
\}_{j=1}^{\nu} \subset \bhhstar{A}^{\otimes l n}$
\begin{equation}
    \label{eqn:coding10}
        \begin{split}
            {\rm tr} \, & T_{n(m+l)*} (\varrho_j) \, M_j \\
            & \geq {\rm tr} \, \tilde{T}_{m+l*}^{\otimes n}
            (\varrho_j) \,
            M_j - \tracenorm{T_{n(m+l)*}  (\varrho_j) -
            \tilde{T}_{m+l*}^{\otimes n} (\varrho_j)}\\
            & \geq {\rm tr} \, \tilde{T}_{m+l*}^{\otimes n} (
            \varrho_j) \, M_j - n \, c^{-m}.
        \end{split}
\end{equation}
For $\varepsilon >0$, choose $n := l$, $m := \varepsilon \, l$ and
$l$ sufficiently large such that Eq.\,(\ref{eqn:coding07}) is
satisfied. We may then conclude from Eq.\,(\ref{eqn:coding10})
that
\begin{equation}
    \label{eqn:coding10a}
        {\rm tr} \, T_{l^2(1+\varepsilon)*} (\varrho_j) \, M_j \,
        > \, 1 - 2 \, \varepsilon
\end{equation}
uniformly in $j$ for sufficiently large $l$, implying that the
product channel random coding scheme leads to asymptotically
vanishing errors for all rates $R < \frac{1}{1+\varepsilon} \,
C_{1}^{} (\tilde{T}_{l})$ and $R < \frac{1}{1+\varepsilon} \,
C_{1}^{p} (\tilde{T}_{l})$, respectively.

We will now show that $(b)$ also holds, with the same substitution
$\varepsilon \mapsto 2 \varepsilon$. To this end, we note that
Devetak's randomization scheme can be slightly modified to include
the output memory state of each block. By this trick we may
guarantee that in an $l$-fold concatenation of blocks of length
$m+l$ each, even the intermediate blocks, for which no coding is
done and the respective outputs are ignored, are (almost)
uncorrelated with Alice's signal states.

Making again use of the error estimate for concatenated channels
and the norm duality Eq.\,(\ref{eqn:duality}), we may then
conclude from Eq.~(\ref{eqn:coding07a}) that
\begin{equation}
    \label{eqn:coding19}
        \begin{split}
            \Big \| & \frac{1}{\nu_{E}} \sum_{k=1}^{\nu_{E}}
            T^{E}_{l(m+l)}
            \left ( \varrho_{jk} \right) \, - \, \Theta   \Big
            \|_{1} \\
            & \leq \Big \| \frac{1}{\nu_{E}} \sum_{k=1}^{\nu_{E}}
            \left [ T^{E}_{l(m+l)} \left ( \varrho_{jk} \right) -
            \tilde{T}^{E \; \otimes l}_{m+l} \left ( \varrho_{jk} \right) \right ]
            \Big \|_{1} \\
            & \qquad + \Big \| \frac{1}{\nu_{E}} \sum_{k=1}^{\nu_{E}}
            \tilde{T}^{E \; \otimes l}_{m+l}
            \left ( \varrho_{jk} \right) - \Theta   \Big
            \|_{1} \\
            & \leq l \, c^{-m} + \varepsilon = l \,
            c^{-\varepsilon l} + \varepsilon \leq 2 \varepsilon
        \end{split}
\end{equation}
for sufficiently large $l$, as advertised. Note that without the
additional randomization over the output memory, the average
mutual information $\frac{1}{l^2} H(A:E)$ between the signal
states and Eve's output states will still be small. This is due to
the fact that in the above coding scheme the intermediate blocks
only constitute a fraction $\varepsilon$ of the total length.
However, this is in general not sufficient to conclude that a norm
estimate such as Eq.~(\ref{eqn:coding19}) holds.

In order to conclude the proof, it only remains to show that $C_1
(\tilde{T}_{l})$ can be bounded from below in terms of $\max_{\{
p_i,\varrho_i \}} \chi(S_{l*}, \{p_i, \varrho_i \})$ for large
$l$, and similarly for the private classical and quantum
capacities.

Applying subadditivity of von Neumann entropy and Fannes'
Inequality Eq.\,(\ref{eqn:fannes}), we see that
\begin{equation}
    \label{eqn:coding11}
        \begin{split}
            \chi & (S_{l*}, \{p_i, \varrho_i \}) \leq
            \chi(T_{l+ \varepsilon l*}, \{p_i, \varrho_i \}) + 2
            \, \varepsilon \, l \, \ld d_B \\
            & \leq \chi(\tilde{T}_{l+ \varepsilon l*}, \{p_i, \varrho_i \}) + 2
            \, \varepsilon \, l \, \ld d_B + \frac{2 \, \ld e}{e} \\
            & \quad \quad + 2 \, l \, (1+\varepsilon) \, \varepsilon \, \ld
            d_B \\
            & \leq l \, (1 + \varepsilon) \, C_1 (\tilde{T}_{l}) + 2
            \, \varepsilon \, l \, \ld d_B + \frac{2 \, \ld e}{e}\\
            & \quad \quad + 2 \, l \, (1+\varepsilon) \, \varepsilon \, \ld
            d_{B}.
        \end{split}
\end{equation}
Since $C_1 (\tilde{T}_{l})$ has been shown to be an achievable
rate for large enough $l$, we may conclude from
Eq.\,(\ref{eqn:coding11}) that
\begin{multline}
    \label{eqn:coding12}
        C_{EE,\mu} (S) \geq \frac{1}{1 + \varepsilon} \Big [ \varlimsup_{l\to\infty}
            \frac{1}{l} \max_{\{ p_i,\varrho_i \}} \chi(S_{l*}, \{p_i, \varrho_i
            \}) \\ - 4 \, \varepsilon \, \ld d_B - 2 \, \varepsilon^2 \, \ld d_B \Big
            ].
\end{multline}
Since $\varepsilon >0$ is arbitrary, Eq.\,(\ref{eqn:coding12})
together with the upper bound in Prop.\,\ref{propo:cbounds}
entails that
\begin{equation}
    \label{eqn:coding13}
        C_{EE,\mu} (S) = \varlimsup_{n\to\infty}
        \frac{1}{n} \max_{\{ p_i,\varrho_i \}} \chi(S_{n*}, \{p_i, \varrho_i
        \}).
\end{equation}
The coding scheme described above uses blocks of length $n_l :=
l^2 (1 + \varepsilon)$. This is a subexponential sequence in the
sense of Remark\,\ref{remark:equivalent}, and we may thus apply
the One-Sequence Theorem \cite{KW04} to conclude that the limit in
Eq.\,(\ref{eqn:coding13}) exists, implying that
Eq.\,(\ref{eqn:coding01}) holds. The rate estimate for the private
classical and quantum capacities is completely analogous.
$\blacksquare$

%%%%%%%%%%%%%%%%%%%%%%%%%%%%%%%%%%%%%%%%%%%%%%%%%%%%%%%%%%%%%%%%%%%%%%%%%%%%%%%%%%
%%%%%%%%%%%%%%%%%%%%%%%%%%%%%%%%%%%%%%%%%%%%%%%%%%%%%%%%%%%%%%%%%%%%%%%%%%%%%%%%%%
%%%%%%%%%%%%%%%%%%%%%%%%%%%%%%%%%%%%%%%%%%%%%%%%%%%%%%%%%%%%%%%%%%%%%%%%%%%%%%%%%%
%%%%%%%%%%%%%%%%%%%%%%%%%%%%%%%%%%%%%%%%%%%%%%%%%%%%%%%%%%%%%%%%%%%%%%%%%%%%%%%%%%

\section{Summary and Outlook}
    \label{sec:sum}

We have presented a general model for quantum channels with
memory, and shown that under mild causality constraints every
quantum process can be thought of as a concatenated memory channel
(plus some memory initializer).

For these memory channels, channel capacities have been introduced
along the lines familiar from the memoryless context, and it has
been demonstrated that different operational setups may lead to
different values of the channel capacity.

While we have concentrated on the classical and quantum channel
capacities proper, it is evident that the theory may be extended
to memory channels assisted by additional resources, such as
entanglement and classical side communication. As seen in
Section~\ref{sec:bounds}, entropic bounds typically depend only on
the amount of information shared by sender and receiver, and not
on the internal structure of the quantum channel linking these
two. Coding theorems for memoryless channels can easily be
extended to forgetful memory channels, as demonstrated in
Section~\ref{sec:coding}. They typically lead to regularized
expressions for the channel capacity, which still require the
solution of optimization problems in Hilbert spaces of
exponentially growing dimensionality. In general, computing
capacities of quantum memory channels is thus at least as
challenging as for memoryless channels, with less hope for
improvements.

A general study of the resulting capacity landscape is still
pending. In particular, we do not yet know under which general
conditions some (or all) of the channel capacities introduced in
Def.~\ref{def:cap} coincide. It may seem reasonable to conjecture
that, as long as the memory system is finite-dimensional, it is
irrelevant for capacity purposes whether Bob or Eve control the
final memory output. While this is almost immediate for the
entropic upper bounds on the channel capacities (cf.
Prop.~\ref{propo:cbounds} and Prop.~\ref{propo:qbounds}), so far
we have not been able to verify this conjecture for the capacities
themselves.

We have demonstrated in Section~\ref{sec:forgetful} that generic
memory channels are forgetful, and in Section~\ref{sec:coding} we
have presented coding theorems for this very important class of
channels. This may seem as if it were possibly to always restrict
one's attention to forgetful channels. However, the capacity of a
memoryless channel is sometimes discontinuous in its parameters.
So while it is always possible to approximate a given
non-forgetful channel by a forgetful channel to arbitrary degree
of accuracy, their capacities may be very different, as the
example given in Section~\ref{sec:forgetful} demonstrates. This
calls for a more detailed analysis of non-forgetful quantum
channels and their capacities.

While we have presented several equivalent criteria for a memory
channel to be forgetful (cf. Section~\ref{sec:forgetful}), we do
not yet have a Structure Theorem to characterize all the
non-forgetful quantum channels, nor do we have a simple test to
decide whether a given memory channel is forgetful.

Apart from some relatively simple model channels, little is known
so far about the channel capacity of general non-forgetful memory
channels. The derivation of coding theorems in this case is likely
to require {\em universal} coding schemes, with encoders and
decoders independent of Eve's choice of the initial memory state.
For the memory channel with a global classical switch (cf.
Section~\ref{sec:examples}), universal coding schemes do exist
\cite{KW05}. However, this is a rather special example of a memory
channel, and the general case remains very much open.

%%%%%%%%%%%%%%%%%%%%%%%%%%%%%%%%%%%%%%%%%%%%%%%%%%%%%%%%%%%%%%%%%%%%%%%%%%%%%%%%%%
%%%%%%%%%%%%%%%%%%%%%%%%%%%%%%%%%%%%%%%%%%%%%%%%%%%%%%%%%%%%%%%%%%%%%%%%%%%%%%%%%%
%%%%%%%%%%%%%%%%%%%%%%%%%%%%%%%%%%%%%%%%%%%%%%%%%%%%%%%%%%%%%%%%%%%%%%%%%%%%%%%%%%
%%%%%%%%%%%%%%%%%%%%%%%%%%%%%%%%%%%%%%%%%%%%%%%%%%%%%%%%%%%%%%%%%%%%%%%%%%%%%%%%%%

\begin{acknowledgments}
    We thank Charles H. Bennett, Igor Devetak, and Andreas Winter for
    fruitful discussions in an enjoyable atmosphere, and Garry Bowen
    for informing us about his work on memory channels with {\em small}
    environments. Alexander S. Holevo, Dirk Schlingemann, and Mario Ziman
    contributed perceptive comments on the manuscript and extremely valuable
    suggestions. Thank yous also go to Sonia Daffer for pointing
    us to \cite{DWC+04}, and to Aram Harrow for sharing his
    insight on compound channels.

    Funding from Deutsche Forschungsgemeinschaft
    (DFG) is gratefully acknowledged.
\end{acknowledgments}

%%%%%%%%%%%%%%%%%%%%%%%%%%%%%%%%%%%%%%%%%%%%%%%%%%%%%%%%%%%%%%%%%%%%%%%%%%%%%%%%%%
%%%%%%%%%%%%%%%%%%%%%%%%%%%%%%%%%%%%%%%%%%%%%%%%%%%%%%%%%%%%%%%%%%%%%%%%%%%%%%%%%%
%%%%%%%%%%%%%%%%%%%%%%%%%%%%%%%%%%%%%%%%%%%%%%%%%%%%%%%%%%%%%%%%%%%%%%%%%%%%%%%%%%
%%%%%%%%%%%%%%%%%%%%%%%%%%%%%%%%%%%%%%%%%%%%%%%%%%%%%%%%%%%%%%%%%%%%%%%%%%%%%%%%%%

\appendix*

\section*{Appendix}
    \label{sec:ql}

\setcounter{equation}{0}

In this section we provide some mathematical background on the
description of infinite-dimensional quantum systems by quasi-local
algebras, and on quantum channels between such algebras. We start
with a quick summary of C$^*$-algebra terminology, and then
concentrate on those aspects which are essential to the proof of
the Structure Theorem in Section~\ref{sec:causal}. For an in-depth
treatment we refer to the texts of Bratteli and Robinson
\cite{BR87}, Ruelle \cite{Rue99}, and Paulsen \cite{Pau02}.

%%%%%%%%%%%%%%%%%%%%%%%%%%%%%%%%%%%%%%%%%%%%%%%%%%%%%%%%%%%%%%%%%%%%%%%%%%%%%%%%%%
%%%%%%%%%%%%%%%%%%%%%%%%%%%%%%%%%%%%%%%%%%%%%%%%%%%%%%%%%%%%%%%%%%%%%%%%%%%%%%%%%%

\subsection{C$^*$-Algebras}
    \label{sec:algebra}

The operations making up the abstract structure of C$^*$-algebras
are inspired by those known from algebras of bounded operators
$\bh$ on a Hilbert space $\hh$. In fact, every such operator
algebra is a C$^*$-algebra, and conversely every abstract
C$^*$-algebra is isomorphic to a norm-closed self-adjoint algebra
of bounded operators on a Hilbert space. More details on this
fundamental structure theorem for C$^*$-algebras will be provided
in Section~\ref{sec:universal} below.\\

A C$^*$-algebra $\A$ is a vector space on the complex numbers $\C$
which is equipped with a product $a \times b \mapsto a \, b$ for
$a, b \in \A$. The product is assumed to be distributive and
associative, but not necessarily commutative. In addition, $\A$
has an {\em adjoint} operation (also called {\em star operation}
or {\em involution}) $\A \ni a \mapsto a^{*} \in \A$. This is {\em
conjugate linear} (or {\em anti-linear}), i.\,e., $(\alpha a +
\beta b)^{*} = \overline{\alpha} a^{*} + \overline{\beta} b^{*}$
for all $a,b \in \A$ and  $\alpha, \beta \in \C$, and has the
properties $a^{**} = a$ and $(ab)^{*} = b^{*} a^{*}$. Physicists
often write $a^{+}$ or $a^{\dagger}$ instead of $a^{*}$.

Besides, there is a {\em norm} $\norm{\cdot}$ on $\A$ which
associates a non-negative number $\norm{a}$ to every $a \in \A$
such that $\norm{a} = 0$ implies $a=0$. With respect to the
algebraic properties of $\A$, the norm satisfies $\norm{\alpha a}
= \abs{\alpha} \, \norm{a}$, the {\em triangle inequality}
$\norm{a+b} \leq \norm{a} + \norm{b}$ and the {\em product
inequality} $\norm{ab} \leq \norm{a} \, \norm{b}$ for all $a, b
\in \A$ and $\alpha \in \C$. In addition, we have $\norm{a^{*}
a^{}} = \norm{a}^2$.\\

An {\em identity} $\1_{\A}$ of a C$^*$-algebra $\A$ is an element
of $\A$ such that $\1_{\A} \, a = a = a \, \1_{\A}$ for all $a \in
\A$. A C$^*$-algebra can have at most one identity. However, not
all algebras come equipped with an identity. The absence of an
identity can complicate the structural analysis, but these
complications can be avoided by embedding $\A$ in a larger algebra
$\tilde{\A}$ which has an identity. Here we will always assume
that $\A$ possesses an identity. Unless the algebra is identically
zero, we then have $\norm{\1_{\A}} = 1$.\\

A {\em state} on the C$^*$-algebra $\A$ is a linear functional
$\omega \mathpunct : \A \rightarrow \C$ which is {\em positive} in
the sense that $\omega (a^{*} a) \geq 0$ for all $a \in \A$ and
normalized such that $\omega (\1_{\A}) = 1$. If $\A = \bhh{A}$ for
some finite-dimensional Hilbert space $\hh_A$, to every state
$\omega$ there exists a unique density operator $\varrho_{\omega}
\in \bhhstar{A}$ such that
\begin{equation}
    \label{eqn:algebra01}
        \omega (a) = \tr{\varrho_{\omega} \, a} \quad \forall \; a
        \in \A.
\end{equation}
For infinite-dimensional systems, there may be states which cannot
be represented as density operators in the sense of
Eq.~(\ref{eqn:algebra01}).\\

The {\em commutant} $\A'$ of a C$^*$-algebra $\A$ is the set of
all operators $a \in \A$ that commute with $\A$, i.\,e.,
\begin{equation}
    \label{eqn:algebra02}
        \A' := \{ a \in \A \, \mid a b = b a \; \forall \; b \in \A
        \}.
\end{equation}
$\A'$ is a sub-algebra of $\A$. If $\A' = \A$, all operators in
$\A$ commute, and the algebra is called {\em Abelian}. These
algebras describe classical systems.

%%%%%%%%%%%%%%%%%%%%%%%%%%%%%%%%%%%%%%%%%%%%%%%%%%%%%%%%%%%%%%%%%%%%%%%%%%%%%%%%%%
%%%%%%%%%%%%%%%%%%%%%%%%%%%%%%%%%%%%%%%%%%%%%%%%%%%%%%%%%%%%%%%%%%%%%%%%%%%%%%%%%%

\subsection{Quasi-Local Algebras}
    \label{sec:qlalgebra}

Quasi-local algebras are adapted to the description of infinitely
extended quantum lattice systems. The framework discussed in this
Section works for any lattice structure in any spatial dimension.
In fact, it does not even require translational invariance and can
be formulated for possibly different quantum (or classical)
systems localized on the nodes of a finite or infinite graph.
However, our interest is in the input and output signals of a
causal automaton, and we may thus restrict our discussion to the
simple case in which the lattice consists of a one-dimensional
spin chain labelled by integers $z \in \Z$. To each site $z \in
\Z$ we assign an isomorphic copy $\A_{z}$ of the observable
algebra $\A$, which in our case is a finite-dimensional
$C^{*}$-algebra $\bhh{A}$ or $\bhh{B}$ of Alice's input and Bob's
output system, respectively. When $\Lambda \subset \Z$ is a finite
subset, we denote by $\A_{\Lambda} := \bigotimes_{z\in\Lambda}
\A_{z}$ the algebra of observables belonging to all sites in
$\Lambda$. Whenever $\Lambda_1 \subset \Lambda_2$, tensoring with
the identity operator $\1_{\A}$ on $\Lambda_2 \setminus \Lambda_1$
will make $\A_{\Lambda_1}$ a sub-algebra of $\A_{\Lambda_2}$. In
the same way the product $a_1 \, a_2$ of operators $a_i \in
\A_{\Lambda_i}$ becomes a well-defined element of $\A_{\Lambda_{1}
\cup \Lambda_{2}}$. Since tensoring with the identity $\1_{\A}$
does not change the norm, this construction yields a normed
algebra of {\em local observables}. Its norm-completion is called
{\em quasi-local algebra}, and will be denoted by
\begin{equation}
    \label{eqn:qlalgebra01}
        \A_{\Z} := \overline{\bigcup_{\Lambda \subset \Z}
        \A_{\Lambda}}.
\end{equation}
Similarly, for infinite subsystems $\Lambda \subset \Z$ we define
$\A_{\Lambda}$ as the closure of the union of all $\A_{\Lambda'}$
for finite $\Lambda' \subset \Lambda$. In particular, by $\A_{-}
:= \A_{(-\infty,0]}$ and $\A_{+} := \A_{[1,\infty)}$ we will
denote the left and right half chain, respectively.\\

The algebra $\A_{\Lambda}$ is interpreted as the algebra of
physical observables for a subsystem localized in the region
$\Lambda \subset \Z$. The quasi-local algebra then corresponds to
the extended algebra of observables on the infinite spin chain
$\Z$.

On the spin chain we introduce a {\em shift operator} $\sigma$ by
setting
\begin{equation}
    \label{eqn:qlalgebra02}
        \sigma \mathpunct : \A_{\Lambda} \rightarrow \A_{\Lambda
        +1} \qquad a \simeq a \otimes \1_{\A} \mapsto \sigma (a) :=
        \1_{\A} \otimes a \simeq a,
\end{equation}
where we have used the notation $\Lambda + 1 := \{ z + 1 \, \mid
\, z \in \Lambda \}.$ The canonical extension of $\sigma$ onto the
quasi-local algebra $\A_{\Z}$ is a $^*$-automorphism on $\A_{\Z}$,
and the integer powers $\{ \sigma^{z} \}_{z\in\Z}$ represent an
action of the translation group $\Z$ by automorphisms on
$\A_{\Z}$.

As explained in Section~\ref{sec:algebra}, a state $\omega$ on the
spin chain is a positive and normalized linear functional on
$\A_{\Z}$. Equivalently, a state $\omega$ is given by a family $\{
\omega_{\Lambda} \}_{\Lambda \subset \Z}$ of density operators on
$\A_{\Lambda}$ for finite $\Lambda \subset \Z$ such that
$\omega(a) = \tr{\omega_{\Lambda} a}$ for $a \in \A_{\Lambda}$.
The local density matrices have to satisfy the consistency
condition that ${\rm tr}_{\Lambda_2 \setminus \Lambda_1}
\omega_{\Lambda_{2}} = \omega_{\Lambda_{1}}$ whenever $\Lambda_1
\subset \Lambda_2$. This equivalence reflects the fact that the
state of the entire spin chain is assumed to be determined by the
expectation values of all observables on finite subsystems
$\Lambda \subset \Z$.

%%%%%%%%%%%%%%%%%%%%%%%%%%%%%%%%%%%%%%%%%%%%%%%%%%%%%%%%%%%%%%%%%%%%%%%%%%%%%%%%%%
%%%%%%%%%%%%%%%%%%%%%%%%%%%%%%%%%%%%%%%%%%%%%%%%%%%%%%%%%%%%%%%%%%%%%%%%%%%%%%%%%%

\subsection{Stinespring's Representation}
    \label{sec:stinespring}

Quantum channels, as introduced in Section~\ref{sec:channel}, are
completely positive and unital maps $S \mathpunct : \B \rightarrow
\A$ between observable algebras $\B$ and $\A$ attributed to
physical systems. In Heisenberg picture language, they describe
how observables (and thus expectation values) transform when the
system under consideration undergoes a free or controlled
evolution.\\

By Stinespring's famous representation theorem \cite{Sti55}, for
every completely positive (not necessarily unital) map $S
\mathpunct : \B \rightarrow \bhh{A}$ we may find a Hilbert space
$\kk$ and an isometry $V \mathpunct : \hh_A \rightarrow \kk$ such
that
\begin{equation}
    \label{eqn:stinespring01}
        S(b) \, = \, V^{*} \, \pi(b) \, V^{} \quad \forall \; b
        \in \B,
\end{equation}
where $\pi \mathpunct : \B \rightarrow \bkk$ is a
$^*$-representation, i.\,e., a linear operator that preserves the
algebraic structure in that $\pi(b_1 \, b_2) = \pi(b_1) \,
\pi(b_2)$ and $\pi(b^{*})^{} = \pi(b^{})^{*}$.

If the output system $\B$ is finite-dimensional, the
representation Eq.~(\ref{eqn:stinespring01}) takes the simpler
form
\begin{equation}
    \label{eqn:stinespring02}
        S(b) \, = \, V^{*} \, (b \otimes \1_{\kk}) \, V^{} \quad \forall \; b
        \in \B
\end{equation}
with the Stinespring isometry $V \mathpunct : \hh_A \rightarrow
\hh_B \otimes \kk$, where $\B = \bhh{B}$ with $\dim \hh_{B} <
\infty$. By means of the duality Eq.~(\ref{eqn:dual}), in
Schr\"odinger picture this form of Stinespring's Theorem gives
rise to the {\em ancilla representation} of the quantum channel
$S_{*}$,
\begin{equation}
    \label{eqn:stinespring03}
        S_{*} (\varrho) = {\rm tr}_{\kk} V^{} (\varrho \otimes
        \varrho_{0}) V^{*} \quad \forall \; \varrho \in
        \bhhstar{A},
\end{equation}
where $\varrho_{0} \in \bkk$ is a so-called {\em ancilla state}.
The Kraus representation Eq.~(\ref{eqn:kraus}) follows from
Eq.~(\ref{eqn:stinespring02}) by introducing a basis
$\{\psi_{i}\}_{i}$ in $\kk$.

A triple $(\kk, \pi, V)$ as obtained in Stinespring's Theorem
Eq.~(\ref{eqn:stinespring01}) is usually called a {\em Stinespring
representation} for the channel $S$. If the closed linear span of
$\pi(\B) V \hh_A$ equals $\kk$, the representation is called {\em
minimal}. Minimal Stinespring representations are unique up to
unitary equivalence, in the following sense: Assume that the
quantum channel $S$ has a minimal Stinespring representation
Eq.~(\ref{eqn:stinespring01}) as well as a further (not
necessarily minimal) one
\begin{equation}
    \label{eqn:stinespring04}
        S(b) \, = \, V_{1}^{*} \, \pi_{1}(b) \, V_{1}^{} \quad \forall \; b
        \in \B
\end{equation}
with another Stinespring isometry $V_{1} \mathpunct : \hh_A
\rightarrow \kk_1$. Since the representation
Eq.~(\ref{eqn:stinespring01}) is assumed to be minimal, we
conclude that $\dim \kk \leq \dim \kk_1$, and the prescription
\begin{equation}
    \label{eqn:stinespring05}
        W (\pi(b) V_{} \psi) \, := \, \pi_{1}(b) V_{1} \psi
\end{equation}
for $b \in \B$ and $\psi \in \hh_{A}$ yields a well-defined
isometry $W \mathpunct : \kk \rightarrow \kk_1$. From the
definition of $W$ we find that the intertwining relation $W \pi =
\pi_{1} W$ holds, implying that $W \pi_{}(b) V_{} = \pi_{1}(b)
V_{1}$ for all $b \in \B$, and thus $W \, V_{} = V_{1}$ by setting
$b = \1_{\B}$. The uniqueness statement plays a central role in
the Structure Theorem for quantum memory channels (cf.
Section~\ref{sec:causal}).

%%%%%%%%%%%%%%%%%%%%%%%%%%%%%%%%%%%%%%%%%%%%%%%%%%%%%%%%%%%%%%%%%%%%%%%%%%%%%%%%%%
%%%%%%%%%%%%%%%%%%%%%%%%%%%%%%%%%%%%%%%%%%%%%%%%%%%%%%%%%%%%%%%%%%%%%%%%%%%%%%%%%%

\subsection{GNS-Representation of Quantum States}
    \label{sec:universal}

A state $\omega \mathpunct : \B \rightarrow \C$, as defined in
Section~\ref{sec:algebra} above, is a unital and positive linear
map. Since the range algebra $\C$ is Abelian, it is even
completely positive (cf. \cite{Pau02}, Th.~3.9), and thus we may
apply Stinespring's Theorem to conclude that $\omega$ can be given
the representation
\begin{equation}
    \label{eqn:universal01}
        \omega(b) := \bra{\Omega} \pi(b) \ket{\Omega} \quad
        \forall \; b \in \B,
\end{equation}
where $\ket{\Omega} := V(1)$. Eq.~(\ref{eqn:universal01}) is
usually called the GNS-representation of quantum states, after
Gelfand and Naimark \cite{GN43}, and Segal \cite{Seg47}.\\
\\
The GNS Theorem can be applied to prove the basic structure
theorem of C$^*$-algebras:
%%%%%%%%%%%%%%%%%%%%%%%%%%%%%%%%%%%%%%%%%%%%%%%%%%%%%%%%%%%%%%%%%%%%%%%%%%%%%%%%%%
\begin{theo}
    \label{theo:fundamental}
        Every {\rm C}$^*$-algebra $\A$ is isomorphic to a norm-closed
        self-adjoint algebra of bounded operators on a Hilbert space.
\end{theo}
%%%%%%%%%%%%%%%%%%%%%%%%%%%%%%%%%%%%%%%%%%%%%%%%%%%%%%%%%%%%%%%%%%%%%%%%%%%%%%%%%%
The idea of the proof is to construct for each state $\omega$ of
$\A$ the corresponding GNS representation $(\kk_{\omega},
\pi_{\omega}, V_{\omega})$, and then to form the so-called {\em
universal representation} by setting
\begin{equation}
    \label{eqn:universal02}
        \kk := \bigoplus_{\omega} \kk_{\omega} \qquad {\rm and}
        \qquad \pi := \bigoplus_{\omega} \pi_{\omega}.
\end{equation}
The existence of sufficiently many states is guaranteed by the
Hahn-Banach extension theorem. The details are spelled out in
Section 2.3 of \cite{BR87}.

%%%%%%%%%%%%%%%%%%%%%%%%%%%%%%%%%%%%%%%%%%%%%%%%%%%%%%%%%%%%%%%%%%%%%%%%%%%%%%%%%%
%%%%%%%%%%%%%%%%%%%%%%%%%%%%%%%%%%%%%%%%%%%%%%%%%%%%%%%%%%%%%%%%%%%%%%%%%%%%%%%%%%
%%%%%%%%%%%%%%%%%%%%%%%%%%%%%%%%%%%%%%%%%%%%%%%%%%%%%%%%%%%%%%%%%%%%%%%%%%%%%%%%%%
%%%%%%%%%%%%%%%%%%%%%%%%%%%%%%%%%%%%%%%%%%%%%%%%%%%%%%%%%%%%%%%%%%%%%%%%%%%%%%%%%%

\end{document}